\documentclass[twocolumn,hidelinks]{aastex631}
\pdfoutput=1 
\usepackage{graphicx}
\usepackage{CJKutf8}
\usepackage{amsmath,amssymb,amstext}
\DeclareFontFamily{U}{mathx}{}
\DeclareFontShape{U}{mathx}{m}{n}{
  <-> mathx10
}{}
\DeclareSymbolFont{mathx}{U}{mathx}{m}{n}
\DeclareMathAccent{\widebar}{0}{mathx}{"73}

\usepackage[utf8]{inputenc}
\usepackage[figure,figure*]{hypcap}
\usepackage[dvipsnames]{xcolor}
\usepackage{bm}
\usepackage{paralist} 
\usepackage{import}
\usepackage{xparse}

\input{macros.sty}
\usepackage{url}

\makeatletter
\let\frontmatter@title@above=\relax
\makeatother

\newcommand{\centrals}[1]{\begingroup\color{Green}\textbf{[Rose et al. (Centrals)]}\endgroup}

\newcommand{\sats}[1]{\begingroup\color{Green}\textbf{[Rose et al. (Satellites)]}\endgroup}

\def\ew{$\widebar{e}_w$ }
\def\kw{\kappa_w }
\def\agn{$\epsilon_{f,\mathrm{high}}$ }

\makeatletter
\DeclareFontFamily{OML}{cmm}{\skewchar\font127 }
\DeclareFontShape{OML}{cmm}{m}{it}{
  <5><6><7><8><9><10><10.95><12><14.4><17.28><20.74><24.88>cmmi10
}{}
\DeclareSymbolFont{cmletters}{OML}{cmm}{m}{it}
\DeclareMathSymbol{\partial}{\mathord}{cmletters}{"40}
\makeatother

\begin{document}

\shorttitle{Speed Distributions in DREAMS}
\shortauthors{Lilie et al.}
\title{The DREAMS Project:\\
Disentangling the Impact of Halo-to-Halo Variance and Baryonic Feedback on\\
Milky Way Dark Matter Speed Distributions}

\correspondingauthor{Ethan Lilie}
\email{elilie@princeton.edu}

\author[0009-0000-8180-9044]{Ethan Lilie}
\affiliation{Department of Physics, Princeton University, Princeton, NJ 08544, USA}


\author[0000-0002-2628-0237]{Jonah C. Rose}
\affiliation{Department of Physics, Princeton University, Princeton, NJ 08544, USA}
\affiliation{Center for Computational Astrophysics, Flatiron Institute, 162 5th Avenue, New York, NY 10010, USA}

\author[0000-0002-8495-8659]{Mariangela Lisanti}
\affiliation{Department of Physics, Princeton University, Princeton, NJ 08544, USA}
\affiliation{Center for Computational Astrophysics, Flatiron Institute, 162 5th Avenue, New York, NY 10010, USA}


\author[0000-0002-8111-9884]{Alex M. Garcia}
\affiliation{Department of Astronomy, University of Virginia, 530 McCormick Road, Charlottesville, VA 22904, USA}
\affiliation{Virginia Institute for Theoretical Astronomy, University of Virginia, Charlottesville, VA 22904, USA}
\affiliation{The NSF-Simons AI Institute for Cosmic Origins, USA}

\author[0000-0002-5653-0786]{Paul Torrey}
\affiliation{Department of Astronomy, University of Virginia, 530 McCormick Road, Charlottesville, VA 22904, USA}
\affiliation{Virginia Institute for Theoretical Astronomy, University of Virginia, Charlottesville, VA 22904, USA}
\affiliation{The NSF-Simons AI Institute for Cosmic Origins, USA}


\author[0000-0003-4004-2451]{Kassidy E. Kollmann}
\affiliation{Department of Physics, Princeton University, Princeton, NJ 08544, USA}

\author[0000-0001-9592-4190]{Jiaxuan Li (\begin{CJK}{UTF8}{gbsn}李嘉轩\end{CJK}\!\!)} 
\affiliation{Department of Astrophysical Sciences, 4 Ivy Lane, Princeton University, Princeton, NJ 08540, USA}

\author[0009-0009-0239-8706]{Olivia Mostow}
\affiliation{The Oskar Klein Centre, Department of Physics, Stockholm University, Albanova University Center, 106 91 Stockholm, Sweden}

\author[0000-0001-7168-8517]{Bonny Y. Wang}
\affiliation{Department of Astronomy and Astrophysics, University of Chicago, Chicago, IL 60637, USA}

\author[0000-0002-7968-2088]{Stephanie O'Neil} 
\affiliation{Department of Physics \& Astronomy, University of Pennsylvania, Philadelphia, PA 19104, USA}
\affiliation{Department of Physics, Princeton University, Princeton, NJ 08544, USA}

\author[0000-0002-6196-823X]{Xuejian Shen} 
\affiliation{Department of Physics and Kavli Institute for Astrophysics and Space Research, Massachusetts Institute of Technology, Cambridge, MA 02139, USA}

\author[0000-0002-0372-3736]{Alyson M. Brooks}
\affiliation{Department of Physics \& Astronomy, Rutgers, the State University of New Jersey, Piscataway, NJ 08854, USA}

\author[0000-0003-0777-4618]{Arya Farahi}
\affiliation{Departments of Statistics and Data Science, University of Texas at Austin, Austin, TX 78757, USA}
\affiliation{The NSF-Simons AI Institute for Cosmic Origins, USA}

\author[0000-0002-3204-1742]{Nitya Kallivayalil}
\affiliation{Department of Astronomy, University of Virginia, 530 McCormick Road,
Charlottesville, VA 22904, USA}

\author[0000-0003-2806-1414]{Lina Necib}
\affiliation{Department of Physics and Kavli Institute for Astrophysics and Space Research, Massachusetts Institute of Technology, Cambridge, MA 02139, USA}
\affiliation{The NSF AI Institute for Artificial Intelligence and Fundamental Interactions, Cambridge, MA 02139, USA}

\author[0000-0002-6021-8760]{Andrew B. Pace}
\affiliation{Department of Astronomy, University of Virginia, 
530 McCormick Road, 
Charlottesville, VA 22904, USA}

\author[0000-0001-8593-7692]{Mark Vogelsberger}
\affiliation{Department of Physics and Kavli Institute for Astrophysics and Space Research, Massachusetts Institute of Technology, Cambridge, MA 02139, USA}
\affiliation{The NSF AI Institute for Artificial Intelligence and Fundamental Interactions, Cambridge, MA 02139, USA}

\date{\today}


\label{firstpage}

\begin{abstract}
Direct detection experiments require information about the local dark matter speed distribution to produce constraints on dark matter candidates, or infer their properties in the event of a discovery. In this paper, we analyze how the uncertainty in the dark matter speed distribution near the Sun is affected by baryonic feedback, halo-to-halo variance, and halo mass. To do so, we harness the statistical power of the new DREAMS Cold Dark Matter simulation suite, which is comprised of 1024 zoom-in Milky Way-mass halos with varied initial conditions as well as cosmological and astrophysical parameters. Applying a normalizing flows emulator to these simulations, we find that the uncertainty in the local DM speed distribution is dominated by halo-to-halo variance and, to a lesser extent, uncertainty in host halo mass. Uncertainties in supernova and black hole feedback (from the IllustrisTNG model in this case) are negligible in comparison. Using the DREAMS suite, we present a state-of-the-art prediction for the DM speed distribution in the Milky Way. Although the Standard Halo Model is contained within the uncertainty of this prediction, individual galaxies may have distributions that differ from it. Lastly, we apply our DREAMS results to the XENON1T experiment and demonstrate that the astrophysical uncertainties are comparable to the experimental ones, solidifying previous results in the literature obtained with a smaller sample of simulated Milky Way-mass halos.
\end{abstract}


\keywords{Hydrodynamical simulations~(767) --- Dark matter distribution~(356) --- Cold dark matter~(265)}
\NewPageAfterKeywords

\section{Introduction}
\label{sec:intro}
A central goal of modern astrophysics and particle physics is to identify the nature of dark matter~(DM). Direct detection experiments represent a key strategy in this effort (see, e.g., \cite{ddreview} for a review). These experiments search for DM particles interacting with Standard Model particles in a terrestrial target. The scattering rates depend on the accurate determination of the phase-space distribution of DM near the Sun, in particular its speed distribution. However, the speed distribution is sensitive to the detailed structure of the Galaxy, the host halo mass, its formation history, the location of the Solar circle, and the treatment of baryonic physics~\citep{newdm1,lmc1,newdm4,newdm3,Anbajagane2022Baryonic,lmc2,lmc3,Staudt_2024,folsom2025darkmattervelocitydistributions}.

The Standard Halo Model~(SHM; see, e.g., \cite{shm1,shm2,shm3}), which was first posited by~\cite{vlsr1}, is often assumed for the local DM speeds. It is a Maxwell--Boltzmann distribution truncated at the Galactic escape speed~\citep{found_shm,found_shm2}. While a distribution like the SHM is expected for a virialized halo, deviations can arise depending on how a halo evolves in a cosmological setting. To this end, cosmological simulations that track the merger history of Milky Way~(MW)-mass halos play an important role in updating our theoretical understanding for the speed distribution.

Initial DM-only simulations of MW-mass halos predicted speed distributions that were typically slower than that expected from the SHM~\citep{earlydm1,earlydm2,earlydm6,earlydm4,earlydm5,earlydm3}. This was largely due to the omission of a baryonic disk, which would otherwise deepen the gravitational potential and increase the DM speeds~\citep{baryons_dm1,baryon1,baryon2,baryon3,baryon4,sales2022baryonicsolutionschallengescosmological,Garcia_to_Appear}.
The next generation of fully hydrodynamic simulations did indeed find that the DM speed distributions shifted to higher values, closer to the local standard-of-rest speed in our Galaxy~\citep{newdm2,newdm6,newdm7,newdm5}. That being said, there was still significant uncertainty comparing the high-speed tails of these distributions with the SHM expectation---especially with regards to the impact of different simulation models for baryonic feedback physics
\citep{newdm2,baryon2,newdm8,newdm6,newdm7,newdm5,newdm1,newdm4,newdm9,newdm3,Lawrence_2022,baryon4}. 

A main challenge in robustly inferring speed distributions from simulations is the limited statistics, with studies often focusing on one or a few simulated galaxies due to computational cost~\citep{newdm2,newdm8,newdm6,newdm7,newdm5,newdm1,newdm4,newdm9,newdm3}. \cite{folsom2025darkmattervelocitydistributions} studied the largest sample to date, using 98 TNG50-1 galaxies \citep{Nelson_2019}. Their study was the first to quantify the degree to which intrinsic halo-to-halo variance---which arises from the unique merger history and evolution of a given halo---affects predictions for the local DM speed distribution.  

Moreover, choosing what is considered a MW analog is non-trivial. Previous works have attempted to account for the unique MW merger history by including a satellite like the Large Magellanic Cloud~(LMC) or a merger like \textit{Gaia} Sausage–Enceladus~(GSE)~\citep{lmc1,lmc2,lmc3,newdm4,folsom2025darkmattervelocitydistributions}. While some find potential impacts on the high-speed tails of the distributions, others find less notable effects.  At this stage, it is difficult to draw definitive conclusions regarding the impact of the merger history given the limited statistics involved in these studies. The likelihood of finding a MW-mass galaxy with both an LMC and GSE is $\lesssim 1\%$~\citep{nadler}, which underscores the computational challenge in generating large-enough samples that match specific details of the MW's evolution.

In addition to matching the merger history, one would also want to ensure that the morphology of the simulated galaxy is similar to our own.  This is particularly important because the MW is compact compared to other galaxies of a similar mass~\citep{compact1, compact2, compact3, compact4}---although see \cite{noncompact}. Therefore, naively setting the solar circle in the simulation to its observed value of 8.3~kpc in the MW~\citep{sun1,sun2} can potentially underestimate the DM speeds. \cite{newdm8} addressed this by only considering simulated galaxies with rotation curves that match our own, which further limited their sample size. In contrast, \cite{Staudt_2024} used an inference procedure to determine properties of the speed distribution consistent with the MW's observed circular velocity 
and \cite{folsom2025darkmattervelocitydistributions} set the solar circle by using an energy-conserving transformation to match the mass enclosed at 8.3~kpc to that of our Galaxy. These approaches are advantageous as they utilize all the simulations in a given suite. 

This paper uses the new DREAMS suite of 1024 MW-mass halos~\citep{Rose_Centrals_to_Appear_A} from the DREAMS project \citep{rose2024introducingdreamsprojectdark} which varies over halo initial conditions and the IllustrisTNG~(TNG; \cite{tng2}) baryonic physics parameters. This is one of the largest simulation suites of high-resolution MW-mass galaxies in Cold Dark Matter~(CDM) and spans a wide range of merger histories and galaxy morphologies. This unprecedented dataset allows us to robustly understand the interplay of different uncertainties on the local DM speed distribution.  
In particular, using a machine-learning-based approach, we isolate the effects of feedback parameters and halo mass on the DM speeds and also quantify the degree to which halo-to-halo variance affects the results. This approach allows us to learn the relationship between MW properties and the DM speed distribution, and we are thus able to make a state-of-the-art prediction for the distribution in our own Galaxy, which can be used in direct detection studies.  

The rest of this paper is organized as follows. Section~\ref{sec:methods} outlines the simulation suite and methods used to calculate the speed distributions. Section~\ref{sec:results} outlines the DREAMS result for the speed distribution specified to our MW galaxy and analyzes how baryonic physics and halo mass variations, in addition to halo-to-halo uncertainty, affect the simulation predictions. Section~\ref{sec:discus} discusses comparisons between DREAMS, TNG50-1, TNG50-2, and FIRE-2 suites and applications to direct detection experiments. We conclude in Section~\ref{sec:concl}. Appendix~\ref{sec:appkde} outlines convergence testing for the procedure used to recover the speed distributions. Appendix~\ref{sec:appemul} details the machine learning architecture used and validation. Appendix~\ref{sec:appsup} includes supplementary figures relevant to the main paper.


\section{Methods}
\label{sec:methods}

This section discusses how to reconstruct a local DM speed distribution from simulation data and how to analyze the interplay of halo-to-halo variance and baryonic parameter variation. We review the CDM DREAMS suite, including the relevant baryonic parameters, cosmological parameters, and galaxy properties. We then introduce a kernel density estimation procedure for recovering the speed distribution, as well as an emulator approach for studying the contributions of halo-to-halo variance, halo mass uncertainty, and baryonic parameter variation. Lastly, we discuss the baryonic parameter space and which regions reproduce results that agree with observations. 

This work uses the \textsc{Subfind}~\citep{subfind} catalogs at $z = 0$  when referring to halos and halo masses. Throughout, the halo mass, $\mhalo$, is defined as the total DM mass gravitationally bound to the halo, excluding orbiting satellite masses.

\subsection{DREAMS Simulations}
\label{sub:dreams}
We use the DREAMS CDM hydrodynamic suite~\citep{Rose_Centrals_to_Appear_A}, which includes 1024 MW-mass zoom-in simulations. This suite uses \textsc{Music}~\citep{music1} to generate initial conditions for the zoom ins at $z=127$, which are then evolved to present day using the moving-mesh code \textsc{Arepo}~\citep{arepo1,arepo2,arepo3}. \textsc{Arepo} models DM gravitational interactions using a TreePM method~\citep{bagla2002} with a comoving spatial softening. Baryons are modeled using a hydrodynamic prescription where they are treated as a fluid solved using finite-volume discretization of ideal magnetohydrodynamics equations.  

The MW-mass zooms are obtained through a multi-step procedure. For each simulation, we first start from a random choice of initial conditions and simulate a low-resolution N-body uniform box. The next step is to simulate an N-body zoom in at intermediate resolution for a random halo in this box with virial mass comparable to the MW.  At this stage, we verify that the target halo is still within the mass range of $(0.50\text{--}2.00)\times 10^{12}\Msun$ and that there is no other halo more massive than $1.0\times 10^{12}\Msun$ within 1~Mpc.  If this is not the case, another halo is selected from the uniform box.  If these criteria are satisfied, then we resimulate the zoom in of this halo at higher resolution and with hydrodynamics. The DM particle mass resolution for the final simulation is $1.8\times \left( {\Omega_{\rm m}}/{0.314} \right) \times10^{6}\Msun$ (where $\Omega_{\rm m}$ is the matter density), the baryon particle mass resolution is $2.8\times10^{5}\Msun$, and the gravitational softening is 0.441~kpc at a redshift of $z= 0$. Note that, while the isolation criterion excludes computationally challenging halos, it also excludes neighbors such as Andromeda~(M31). One should therefore keep in mind that the hosts analyzed in this work are not exact MW analogs and may have different environments and/or formation histories.

The DREAMS suite uses the TNG galaxy-formation and feedback model~\citep{tng2}. The TNG model includes prescriptions for star formation, supernovae feedback, active galactic nuclei~(AGN) feedback, black hole formation, DM, gravity, and other astrophysical processes.
The DREAMS suite varies three parameters of the TNG model that are related to baryonic feedback. The first two are related to supernova feedback through the mass-loading factor 
\begin{equation}
    \label{eq:mass_loading}
    \eta_w = \frac{2}{v_w^2} e_{w} \left(1 - \tau_w \right) \, ,
\end{equation}
where $e_{w}$ is the specific energy available for generating winds, $v_w$ is the wind speed, and $\tau_w$ is the fraction of thermal energy released.  The DREAMS simulations specifically vary the dimensionless specific energy scaling for supernovae winds, defined through
\begin{align}
    e_w &= \left( \frac{\widebar{e}_w}{\ 10^{-51} \ \mathrm{erg}^{-1}\Msun} \right) f(Z) \, N_\mathrm{SN} \, ,
\end{align}
where $f(Z)$ is a function that reduces the available energy as a function of metallicity~($Z$), and $N_\mathrm{SN}$ is the number of Type II supernovae per formed stellar mass. \ew is varied logarithmically between \ew~$\in~[0.9, 14.4]$ and $\tau_{w}$ is set to the TNG fiducial value of 0.1. The fiducial value for \ew in the TNG model is 3.6. 

The speed of the galactic winds created by SN feedback is given by
\begin{align}
v_w = \mathrm{max} \left[\kw  \, \sigma_{\scriptscriptstyle{\rm DM}} \left(\frac{H_0}{H(z)} \right)^{1/3}, \, v_{w,\mathrm{min}} \right] \, ,
\end{align}
where $\kw$ is a dimensionless normalization factor,  $\sigma_{\scriptscriptstyle{\rm DM}}$ is the DM velocity dispersion around the supernova, $H(z)$ is the Hubble parameter, and $v_{w,\mathrm{min}}$ is the minimum wind speed (set to $350~\kms$ ). 
$\kw$ is varied logarithmically between $\kw \in [3.7, 14.8]$; its fiducial value in the TNG model is 7.4. The two parameters \ew and $\kw$ strongly influence the stellar mass of a galaxy~\citep{tng2,camels,Rose_Centrals_to_Appear_A,Garcia_to_Appear}. Therefore, lower values of these two parameters lead to higher stellar masses, which deepens the potential well compared to galaxies with higher values of \ew and $\kw$. 

AGN feedback is scaled using the parameter \agn\!\!\!, which corresponds to the fraction of energy transferred to nearby gas. The strength of AGN feedback can impact the DM halos around galaxies~\citep{2022Anbajagane}. The TNG model contains a low-accretion state and a high-accretion state, however MW-mass galaxies spend most (or all) of their lives below the stellar mass where low-accretion feedback dominates~\citep{weinberger2017}. Therefore, the DREAMS suite only varies over the high-accretion mode. In this state, the feedback energy released from accretion is given by 
\begin{equation}
    \dot{E}_{\rm AGN} = {\epsilon_{f,\mathrm{high}}} \, \epsilon_r \, \dot{M}_{\scriptscriptstyle{\rm BH}} \,  c^2 \, ,
\end{equation}
where $\epsilon_r$ is the radiative efficiency, $\dot{M}_{\scriptscriptstyle{\rm BH}}$ is the black hole accretion rate, and $c$ is the speed of light. 
\agn is varied logarithmically in the range $\epsilon_{f,\mathrm{high}} \in [0.025, 0.4]$. 
Its TNG fiducial value is 0.1. 

The DREAMS CDM suite also varies over some cosmologies, with $\Omega_{\rm m}$, the total matter density, and $\sigma_{8}$, the amplitude of matter fluctuations, being uniformly sampled across all 1024 simulations within $[0.274 ,0.354]$ and $[0.780, 0.888]$, respectively. These ranges correspond to two standard deviations on the Planck 2013 temperature data \citep{Planck2013}. This corresponds to a range of seven standard deviations below the mean to eight standard deviations above the mean for $\Omega_{\rm m}$ 
and five standard deviations below the mean and thirteen standard deviations above for $\sigma_{8}$, using the Planck 2018 full-mission results~\citep{2020Planck}. The other cosmological parameters, $H_{0} = 69.1\kms \rm{\ Mpc^{-1}}$ and $\Omega_{\rm b} = 0.046$, are fixed for each simulation and are consistent with Planck 2016 values~\citep{2016Planck}.

All simulation parameters are varied simultaneously for the 1024 simulations according to a Sobol sequence~\citep{sobol}. The range of supernovae feedback parameters are consistent with observations of star formation rate densities~\citep{vogel1,vogel0} when individual parameters are varied, excluding the high end of \ew and $\kw$ which are not consistent with observations. See \cite{rose2024introducingdreamsprojectdark} for more details on the choice of parameter variation. 


\subsection{Dark Matter Speed Distributions}
\label{sub:DM_vd}

Previous works studying the local DM speed distribution often take all DM particles within an annulus centered at the solar circle and then create a histogram of those speeds~\citep{newdm2,newdm8,newdm6,newdm7,newdm5,newdm1,newdm4,newdm9,newdm3,Staudt_2024,folsom2025darkmattervelocitydistributions}. The width and height of the annulus must be small enough to adequately approximate the speed distribution near the solar circle, but large enough to contain a sufficient number of DM simulation particles. These works typically set the solar circle to its corresponding value in the MW, $\sim 8.3$~kpc~\citep{sun1, sun2}. In contrast, \cite{Staudt_2024} and \cite{folsom2025darkmattervelocitydistributions} use interpolation and scaling methods to set the location of the solar circle in a more principled manner, but still use the annulus method at this updated location. Here, we implement a novel approach using kernel density estimation~(KDE), which has the advantage of minimizing the spatial scale over which the speed distribution is determined, sidestepping the issue of setting the size of the annulus.

The KDE procedure treats the simulation particles as realizations of the underlying distribution and weights them using a smoothing function based on their distance from the location where the speed distribution is being calculated. Consider dividing the particle speeds into $i$ separate bins, each containing $N_i$ simulation particles. Then the  probability density of speed $v_{i}$ at some  location $\boldsymbol{r}_{0}$ is given by 
\begin{equation}
    \hat{f}(v_{i}) = \frac{1}{N}\sum_{k=1}^{N_i}W(|\boldsymbol{r}_i^{k}-\boldsymbol{r}_{0}|,h)\, ,
\end{equation}
where $W(|\boldsymbol{r}_i^{k}-\boldsymbol{r}_{0}|,h)$ is the kernel function with width $h$, and $\boldsymbol{r}_i^{k}$ is the position of the $k^{\rm th}$ particle in the bin. The probability density is normalized by 
\begin{equation}
N = a\sum_{i=1}^{N_{\rm tot}}\sum_{k=1}^{N_i} W(|\boldsymbol{r}_i^{k}-\boldsymbol{r}_{0}|,h)\, ,
\end{equation}
where $a$ is the bin width and $N_{\rm tot}$ is the total number of bins. We use 60 bins in a range of speeds from 0 to 650$\kms$, so $a = 10.83~\kms$. The kernel function is a cubic spline:
\begin{equation} 
\label{eq:kernel}
W(r,h) = \dfrac{8}{\pi h^{3}}\begin{cases}
1 - 6\left(\dfrac{r}{h}\right)^2 + 6\left(\dfrac{r}{h}\right)^3 & 0 \leq r/h \leq 1/2 \\ 
2\left(1 - \left(\dfrac{r}{h}\right)\right)^{3} & 1/2 < r/h \leq 1 \\
0 & 1 < r/h \, ,
\end{cases} \end{equation}
where $r$ is the distance from the particle to the sample point, and $h$ is the width of the kernel~\citep{Monaghan_1992}. While there is no one uniquely justified choice for the kernel function, the cubic spline allows particles that are spatially farther apart to be down weighted and introduces a scale $h$ for which particles outside this distance hold no weight. This allows one to reconstruct a speed distribution local to the sample point. 

The procedure described above calculates $\hat{f}(v)$ for a specific spatial location.  To obtain the speed distribution across the entire solar circle, we repeat this process for many points along it. Therefore, in addition to $h$, there are two other free parameters to set: the number of points at which the speed distribution is determined~($N_\mathrm{samples}$) and the number of particles used to estimate the speed distribution at each sample point~($N_\mathrm{near}$).  Our baseline analyses are performed with $h=2$~kpc, $N_{\rm samples} = 200$, and $N_{\rm near} = 50$.  After performing the KDE procedure at these various points, we combine the results to infer the speed distribution for the entire solar circle.
We set the solar circle to 8.3~kpc (see Section~\ref{sub:restrict_mw} on the justification for this choice and our approach to ``MW analogs''), but find that any radius from 8.0--8.3~kpc does not significantly change the results.

For the 1024 galaxies in the DREAMS suite, this procedure gives effective KDE radii, the average distance from the farthest particle at each sample point, in the range of 1--2~kpc around a radius of 8.3~kpc. These distances are comparable to the range of annulus sizes used for previous works~\citep{Staudt_2024,folsom2025darkmattervelocitydistributions}, demonstrating that the spatial scales are comparable. However, because this number depends on the 50 nearest particles, this will be resolution-dependent, and thus lower for higher-resolution simulations. The effective KDE radius is different for each galaxy because different galaxy morphologies correspond to different neighbor distances. 

Our core results and conclusions are unchanged based on the choice of $h$, $N_\mathrm{samples}$, or $N_\mathrm{near}$ as long as $h$ is larger than $\sim 3\times$~the softening length of the simulations and $N_\mathrm{near} > 2$. (See Appendix \ref{sec:appkde} for more details and cross-validation.) The KDE approach agrees with the results of the annulus method where the particle speeds within the annulus (with height of $2~\rm{kpc}$ and radius from $7.3\text{--}9.3~\rm{kpc}$) are binned in a histogram. The KDE approach is in principle more robust in estimating the speed distribution, however. The annulus acts effectively as a top-hat kernel with a radius the size of the kernel width. 

\subsection{Normalizing Flows Emulator}
\label{sub:nf}
The 1024 DREAMS galaxies are unprecedented in scope, providing a first opportunity at adequately sampling uncertainties from simulation parameters. Analyzing how these parameters affect the local speed distribution is however challenging due to the high-dimensional nature of this space, which is further complicated by the fact that each host has a unique formation history. An additional challenge arises from the fact that the parameters are discretely sampled and one would ideally like to interpolate between their values to solidify these correlations.

To this end, we employ a normalizing flows emulator in this work. The emulator acts as an interpolator for this large parameter space, taking as input a selection of simulation parameters and outputting a binned speed distribution. As we aim to understand how the baryonic parameters impact the speed distributions, we include $\widebar{e}_w$, $\kw$, and \agn\!\! as input parameters. We exclude the cosmological parameters, $\Omega_{\rm m}$ and $\sigma_{8}$, as we find when these are included they have no effect on the resultant speed distributions. However, because effects of baryonic physics can be degenerate with different halo masses, we include $\mhalo$ as an input parameter as well.
To infer the parameters between those sampled by the DREAMS galaxies, we use the normalizing flows procedure described in \cite{nguyen2024dreamsmadeemulatingsatellite} and \cite{Cuesta-Lazaro:2023zuk}. This starts from some probability distribution $p(\boldsymbol{y})$, for parameters of interest $\boldsymbol{y}$ (e.g., $\hat{f}(v_{i})$, $\mstar$, $\mlsr$). The parameters $\boldsymbol{y}$ then go through $T$ discrete invertible transformations: 
\begin{equation}  \boldsymbol{u}_T = g^{-1}_{T} \circ g^{-1}_{T-1}\ldots \circ g^{-1}_{1}(\boldsymbol{y},\widebar{e}_{w},\kw,\epsilon_{f,\mathrm{high}},\mhalo) \,,
\end{equation}
where $u_{T}$ is the final transformed parameter and each $g_{n}$ is invertible and depends on \ew\!\!, $\kw$, \agn\!\!, and $\mhalo$ such that 
\begin{equation} 
p(\boldsymbol{y}) = p(\boldsymbol{u}_T)\prod_{n=1}^{T}\left|\det\frac{\partial g^{-1}_{n}}{\partial \boldsymbol{u}_{n-1}} \right|\, ,
\end{equation}
where $\boldsymbol{u}_{n} = g^{-1}_{n}\circ \ldots \circ g_{1}^{-1}(\boldsymbol{y},\widebar{e}_{w},\kw,\epsilon_{f,\mathrm{high}},\mhalo)$, and thus $\boldsymbol{u}_{0} = \boldsymbol{y}$. By minimizing the log likelihood, 
\begin{equation} \mathcal{L}_{\text{flows}}(\boldsymbol{y}) = -\log p(\boldsymbol{u}_T)-\sum_{n=1}^{T}\log \left|\det\frac{\partial g^{-1}_{n}}{\partial \boldsymbol{u}_{n-1}} \right|\,, 
\end{equation}
one finds the transformations $g_{n}$. These functions are parameterized as neural networks with the weights and biases obtained through the minimization of the log likelihood. The base distribution $p(\boldsymbol{u}_{T})$ is set to be Gaussian with mean and standard deviation of the sample data. 

With the relationship between $p(\boldsymbol{u}_{T})$ and $p(\boldsymbol{y})$ understood, the process can now be reversed. We can input values of \ew\!\!, $\kw$, \agn\!\!, and $\mhalo$ and start from $p(\boldsymbol{u}_{T})$ to generate a distribution of $p(\boldsymbol{y})$. The emulator acts as an interpolator for values of \ew\!\!, $\kw$, \agn\!\!, and $\mhalo$ within the DREAMS parameter space. Note that the distribution $p(\boldsymbol{y})$ is not the distribution speed $\hat{f}(v)$. It is the distribution of parameters $\boldsymbol{y}$ across the 1024 DREAMS simulations. For example, we have a distribution of $\hat{f}(v_{i})$ for each bin $i$ and a distribution of $\mstar$. 

The emulator is conditioned on \ew\!\!, $\kw$, \agn\!\!, and $\mhalo$. Because the DM speed distributions are not dependent on the cosmological parameters, $\Omega_{\rm m}$ and $\sigma_{8}$, they are excluded as inputs to the emulator. The emulator targets the height of the speed distribution histograms in 60 bins from $0\text{--}650~\kms$.\footnote{In testing, we find that there are virtually no DM particles with speeds higher than $650\kms$ in the DREAMS simulations.}
The emulator can also predict MW properties, in particular we use the stellar mass of the MW, $M_\star$, and the mass enclosed at the solar circle, $M_{\mathrm{LSR}}$. Thus, the emulator output provides a binned DM speed distribution, as well as other MW properties, for any set of \ew\!\!, $\kw$, \agn\!\!, and $\mhalo$.

To ensure that the emulator is producing accurate results, we attempt to recreate the DREAMS suite using the emulator. We generate 1024 samples using the emulator where we input for each parameter the exact parameters (\ew\!\!, $\kw$, \agn\!\!, and $\mhalo$) for the 1024 DREAMS simulations and compare the resultant speed distributions to the DREAMS simulation output (see Appendix~\ref{sec:appemul} and Figure~\ref{fig:emul}). We find that the emulator is able to accurately reproduce the distributions up to the 84th and 16th percentile, with minor disagreement in the extrema of the 98th and 2nd percentiles. To quantify this disagreement, we use the Earth Mover's Distance~(EMD) which can informally be thought of as the ``cost'' or ``work'' to transform one distribution into another~\citep{emd1}. The median of the emulator result and the simulation output has an EMD of $0.7\kms$---for the 2nd percentiles, the EMD is $4.9\kms$, the 16th percentiles $2.3\kms$, the 84th percentiles $1.9\kms$, and the 98th percentiles $7.8\kms$.
This suggests that the emulator learns the relationship between the speed distributions and the baryonic feedback parameters within a few $\kms$ (smaller than the width of our bins).
See Appendix \ref{sec:appemul} for more details on emulator validation. 

\subsection{Parameter-Space Weighting Scheme}
\label{sub:weights}

The DREAMS suite varies over a number of simulation parameters and regions of this space can produce galaxies that are inconsistent with observations, thus overestimating this theoretical uncertainty. We account for this using the weighting method described in \cite{Rose_Centrals_to_Appear_A}, where regions of parameter space are weighted with a preference for those close to observed scaling relations. This allows us to down-weight parts of the parameter space of \ew\!\!, $\kw$, and \agn which do not reproduce the observed scaling relation for populations of galaxies in our Universe. The scaling relation we use is the stellar mass--halo mass relation~\citep{SagaV}. This scaling relation is fit with a piecewise linear function $g(X)$, where $X$ is halo mass and $g$ is piecewise such that the slopes and intercepts best fit the observed stellar mass--halo mass relation.

In practice, the \ew\!\!, $\kw$, and \agn parameter range is sectioned into a $30\times30\times30$ grid. For $2.7\times 10^4$ astrophysical parameter combinations, each labeled $\theta_{j}$ for the $j^{\rm th}$ bin, $10^3$ halo samples are generated using a normalizing flows emulator~\citep{nguyen2024dreamsmadeemulatingsatellite}. Each sample, labeled $m$, consists of a pair of $(X^{\rm{em}}_{j,m},Y^{\rm{em}}_{j,m})$, which correspond to log halo mass and log stellar mass,  respectively. Each parameter bin is associated with a residual 
\begin{equation}
    R_{\theta_{j}} = \frac{1}{N_{j}}\sum_{m=1}^{N_{j}}(Y^{\rm{em}}_{j,m}-g(X^{\rm{em}}_{j,m})) \, ,
\end{equation}
where $N_{j} = 10^3$ is the total number of samples per bin.  The residual is then converted into a weight, $w_{j}$, following 
\begin{equation}
     w_{j} \ \propto \ \exp \left({\frac{-R_{\theta_j}^{2}}{2\tau^{2}}}\right)~\, ,
\end{equation}
where $\tau$ is a hyperparameter set to 0.2 dex. This parameter controls the degree to which we penalize deviations from the observed scaling relation, acting as the uncertainty. The value of 0.2 dex is set based on the uncertainty in the observed mean relation~\citep{2019Behroozi}. See \cite{Rose_Centrals_to_Appear_A} for a more detailed discussion of this. As our main results are not particularly sensitive to the weighting scheme overall, $\tau$ in the range of 0.2--0.3 dex does not change the results of this paper. The weights are normalized such that the sum over all bins is 1. 

This approach does not reproduce the scaling relations exactly, but instead returns a preference for regions of parameter space which reproduce galaxies close to them. Specifically, simultaneous extreme values of \ew and $\kw$ are down-weighted the most as they produce galaxies with too-low stellar mass or too-high stellar mass. 

The weights are applied in Figure~\ref{fig:milkyway}, Figure~\ref{fig:fire}, Figure~\ref{fig:dd}, and Figure~\ref{fig:emd}. This is done by emulating many speed distributions for different values of \ew\!\!, $\kw$, and \agn\!\!. Then, we take the median of these distributions and the weight of each is given by $w_{j}$. Figure~\ref{fig:weights} compares the main result of the paper, Figure~\ref{fig:milkyway}, with and without weights. 

\subsection{Restricting to the Milky Way}
\label{sub:restrict_mw}

\begin{figure*}
  \centering
  \includegraphics[width=\textwidth]{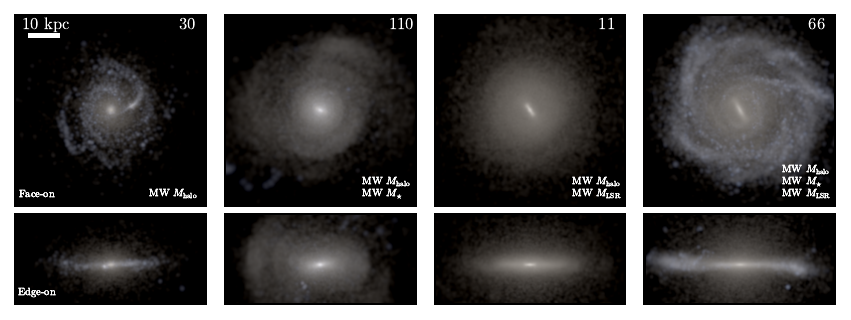}
  \caption{
Three-dimensional stellar light reconstructions of the central galaxies from four of the DREAMS CDM zoom-in simulations. The images are made from the $i, r, g$ Sloan broadband filters. Each image shows a $70 \times 70 \times15$ kpc$^{3}$ volume. The top panels show face-on images and the bottom panels show edge-on images. The DREAMS box number is labeled in the upper-right corner of each top panel. The leftmost image is a galaxy with $\mstar$ and $\mlsr$ less than that of the MW, the middle-left image is a galaxy with $\mstar$ within the uncertainty of the MW but not $\mlsr$, the middle-right image is a galaxy with $\mlsr$ witin the MW uncertainty but not $\mstar$, and the rightmost image shows a galaxy with $\mstar$ and $\mlsr$ within the uncertainty of our Galaxy. Galaxies such as the rightmost one are what we use for the analysis in Figure~\ref{fig:milkyway}.}
  \label{fig:galaxy_images}
\end{figure*}

Even after down-weighting parts of the simulation parameter space, the DREAMS galaxies exhibit a wide range of galactic properties and formation histories---see \cite{Rose_Centrals_to_Appear_A}. For example, they have different stellar masses, disk scale lengths, and disk scale heights, as well as differing enclosed masses at the solar radius. Additionally, our Galaxy is uniquely compact for MW-mass systems~\citep{compact1, compact2, compact3, compact4}, which has important consequences for the local DM speeds~\citep{folsom2025darkmattervelocitydistributions}. The DREAMS suite contains galaxies ranging in stellar mass from $\log \mstar/\Msunt = 8.54\text{--}11.32$ with enclosed masses (at the solar radius) from $\log M_{\rm LSR}/\Msunt = 10.17\text{--}11.44$. The corresponding values for our own Galaxy are $\log \mstar/\mathrm{M}_{\odot}  = 10.7\pm 0.09$ and $\log M_{\mathrm{LSR}}/\mathrm{M}_{\odot} = 11.03 \pm 0.055$~\citep{Bland_Hawthorn_2016}. In the DREAMS suite, $19\%$ of the simulations have $\mstar$ in this range,\footnote{This fraction differs slightly from that quoted in \cite{Rose_Centrals_to_Appear_A} as they only refer to the subset of disk galaxies in DREAMS.} $15\%$ have $\mlsr$ in this range, and  $5\%$ have both in the observed range. 

Figure~\ref{fig:galaxy_images} shows four example galaxies from the DREAMS CDM suite to give a sense of the different morphologies. The leftmost panel is a galaxy with $\mstar$ and $\mlsr$ less than the MW values ($\log \mstar/\Msunt = 10.09$ and $\log \mlsr/\Msunt = 10.70$), the middle-left panel is a galaxy with $\mstar$ within the MW uncertainty but not $\mlsr$ ($\log \mstar/\Msunt = 10.66$ and $\log \mlsr/\Msunt = 10.86$), the middle-right panel is a galaxy with $\mstar$ above, but $\mlsr$ within, the MW uncertainty ($\log \mstar/\Msunt = 10.84$ and $\log \mlsr/\Msunt = 11.05$), and the rightmost panel is a galaxy with $\mstar$ and $\mlsr$ within the MW uncertainty ($\log \mstar/\Msunt = 10.72$ and $\log \mlsr/\Msunt = 10.96$). In other words, the leftmost galaxy does not have enough stars, the left-middle galaxy is not compact enough, the middle-right galaxy is too massive, and the rightmost galaxy is the most comparable to our MW.

To get a speed distribution relevant for direct detection experiments, we restrict to values of $\mstar$ within the MW uncertainty. However, because the MW is compact compared to other galaxies of similar mass, only specifying stellar mass is not sufficient. Different approaches have been used in the literature to account for this. For example, \cite{Staudt_2024} used a fitting procedure to infer the circular velocity at the solar circle, while \cite{folsom2025darkmattervelocitydistributions} rescaled all galaxies to have the same $M_{\mathrm{LSR}}$ at the solar circle. We take a different approach in this work, using an emulator to learn the connection between the speed distribution and the two mass parameters, $\mstar$ and $\mlsr$. We have verified that restricting other parameters such as disk scale length and disk scale height does not change the speed distribution (whether or not we are also restricting on $\mlsr$). 

Beyond restricting to $M_{\star}$ and $M_{\rm LSR}$, we do not make any further requirements on the MW analogs, such as on their formation history.  As  explored in previous work, the LMC and GSE merger can have potential effects on the local DM speed distribution~\citep{lmc1,newdm4,lmc2,lmc3,folsom2025darkmattervelocitydistributions}.  While principled to include these, the statistics on these analogs are limited. The number of LMC analogs in the DREAMS suite is likely low, as \cite{nadler} found the fraction of MW-mass galaxies with LMC analogs is $\sim 2\%$. The number of GSE analogs in the DREAMS CDM suite is $\sim 15$, as discussed in~\cite{Rose_Centrals_to_Appear_A}. We leave an exploration of these effects to future work. 

\subsection{Contributions to Uncertainty}

Throughout this work, we probe three main contributions to the local DM speed distribution. This includes uncertainty from the simulation parameter variation, halo mass variation, and lastly the intrinsic population variance at fixed halo mass and fixed parameters. This last quantity we consider as ``halo-to-halo variance''. This variance can be reduced by conditioning on other variables, such as $\mstar$ and $\mlsr$, as done in Figure~\ref{fig:milkyway}. The uncertainty bands shown in Figure~\ref{fig:milkyway} should be understood as a combination of simulation  parameter variation (dominated by baryonic feedback), halo mass variation, and halo-to-halo variance. Each uncertainty band in the bottom panel of Figure~\ref{fig:mhalo} is the halo-to-halo variance at a unique halo mass. Whereas for Figure~\ref{fig:snfeedback}, each uncertainty band at fixed values of \ew (or $\kw$) is the halo-to-halo variance for different baryonic parameters at $\log \mhalo/\Msunt = 12$.

\section{Results}
\label{sec:results}

\begin{figure}
  \centering
  \includegraphics{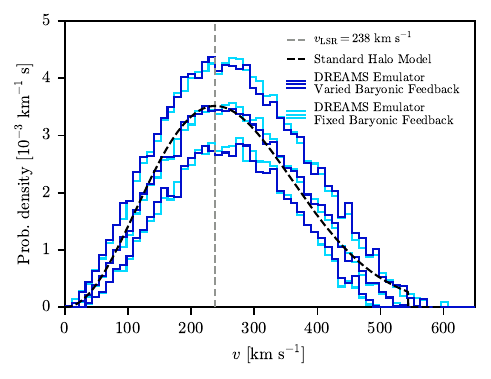}
  \caption{
Median local DM speed distribution, and 16th--84th percentile band, for the emulated DREAMS dataset with Varied~(dark blue) and Fixed~(light blue) baryonic feedback parameters. These distributions are local to the solar radius at 8.3 kpc, and this follows for all other plots shown in this work. For the Fixed dataset, the parameters are set to their TNG fiducial values. The baryonic feedback parameters are marginalized over for the Varied band, while the halo mass is marginalized over for both Varied and Fixed.  For each case, we restrict to systems with $ 10.60 \leq \log\mstar/\Msunt \leq 10.78$ and $  10.975 \leq \log\mlsr/\Msunt \leq  11.085$, consistent with observations for our Galaxy~\citep{Bland_Hawthorn_2016}. For context, the dashed-gray line denotes $v_{\rm LSR} = 238\kms$, while the dashed-black line denotes the Standard Halo Model~(SHM). The results for Fixed and Varied baryonic physics are consistent with each other, suggesting that baryonic feedback uncertainty is subdominant to halo-to-halo variance and halo mass uncertainty (see also Figure~\ref{fig:emul_val_2}). Additionally, the SHM falls within the ensemble of distributions found in the emulated datasets, suggesting that it is a good proxy even if it may not exactly reproduce individual galaxy distributions. }
  \label{fig:milkyway}
\end{figure}

The key results of the paper are obtained by using an emulator that takes as input $\widebar{e}_{w}, \kw, \epsilon_{f,\mathrm{high}},$ and $\mhalo$.  It then outputs $M_*$, $M_{\rm LSR}$, and the binned speed distribution for the DM at the solar circle~(galactocentric radius of 8.3~kpc) using $10^5$ samples. The dark blue lines in Figure~\ref{fig:milkyway} show the 16-50-84th percentiles of the speed distribution obtained after selecting galaxies with $ 10.60 \leq \log\mstar/\Msunt \leq 10.78$ and $  10.975 \leq \log\mlsr/\Msunt \leq  11.085$, consistent with the expectations for our Galaxy~\citep{Bland_Hawthorn_2016}. The baryonic parameters and halo mass are randomly sampled from a uniform distribution in log space across the full ranges varied over in the DREAMS simulations.\footnote{The speed distribution is weighted using the scheme described in Section~\ref{sub:weights}. The corresponding result without the weighting scheme is shown in Figure~\ref{fig:weights}; the overall conclusions are unchanged.} Selecting on emulated galaxies with $\mlsr$ similar to our own should give speed distributions consistent with the observed local standard-of-rest speed of $v_{\mathrm{LSR}} = 238\pm15\kms$~\citep{vlsr1,vlsr2,Bland_Hawthorn_2016}.  Indeed, the emulated speed distribution peaks near $238\kms$, which is indicated by the vertical dashed-gray line.

The light blue lines in Figure~\ref{fig:milkyway} show the corresponding speed distribution when the baryonic feedback parameters are fixed to the fiducial TNG values and halo mass is varied over as before.
There is good agreement between the emulator with Varied and Fixed baryonic physics, which suggests that the uncertainty is dominated by halo-to-halo variance and halo-mass uncertainty. In particular, the Varied and Fixed datasets have a median $v_{\rm LSR}$ of  $245^{+38}_{-39}\kms$ and $v_{\rm LSR} = 258^{+34}_{-38}\kms$, respectively. 

Next, we explore the dependence of the speed distribution on specific feedback parameters (see also Figure~\ref{fig:emul_val_2}). Marginalizing over the other emulator inputs and excluding $\mstar$ and $\mlsr$ constraints, the median $v_{\rm LSR}$ varies by $\sim 57\kms$ from the lowest to highest \ew values, while the scatter is $\sim 100\kms$ across this range. Similarly, $v_{\rm LSR}$ varies by $\sim 71\kms$ from lowest to highest $\kw$ values, while the scatter is $\sim 100\kms$ across this range. This uncertainty is dominated by halo-to-halo variance, as when we fix halo mass, the overall scatter decreases by only $\sim 20\%$.

In addition to the properties of the peak of the distribution, we also study its high-speed tail. To this end, we define $v_{99}$ as the $99^{\rm th}$ percentile of the speed distribution. For the Varied and Fixed datasets, $v_{99}$ is essentially indistinguishable with  $498^{+32}_{-32}\kms$ for both. Additionally, the median $v_{99}$ changes by $\sim 45~(83)\kms$ going from low to high \ew$\!\!~(\kw)$, when marginalizing over the other inputs and excluding $\mstar$ and $\mlsr$ constraints. In comparison, the scatter in the distributions is typically greater than $\sim 110\kms$ over the same range. The uncertainty in $v_{99}$ is more sensitive to $\mhalo$ than $v_{\rm LSR}$, as fixing the halo mass decreases the overall scatter by $\sim 30\%$.

The black dashed line in Figure~\ref{fig:milkyway} shows the SHM, a Maxwellian peaked at $v_{\mathrm{LSR}} = 238\kms$ and cut off at the escape speed $v_{\mathrm{esc}} = 544\kms$.  The SHM is consistent with the emulated DREAMS result, falling within the 16--84$^{\rm th}$ percentile range across all speeds.  This confirms that the SHM is a good proxy for the ensemble behavior of speed distributions in MW-mass galaxies, as discussed in \cite{folsom2025darkmattervelocitydistributions}, even if the distribution of an individual galaxy may differ from it.  
While the SHM is still encompassed within the spread of the Varied and Fixed bands above $\sim 450\kms$, the median of the emulated distributions does fall off faster. For the SHM $v_{99}$ is $514\kms$, which is greater than the corresponding values for the Varied and Fixed datasets (see also Figure~\ref{fig:emd}). These small offsets result in an EMD between the emulated distribution and the SHM that is $11.2^{+9}_{-5}\kms$ and $11.5^{+9}_{-5}\kms$ for the Varied and Fixed datasets, respectively.  Figure~\ref{fig:emd} provides the full EMD distribution for the emulated DREAMS datasets assuming Varied and Fixed baryonic physics.  

The next two subsections analyze the dependence of the DM speed distribution with galaxy properties---specifically halo mass and the baryonic feedback parameters---in greater detail. 

\subsection{Halo Mass}
\label{sub:masses}

\begin{figure}
\centering
\includegraphics{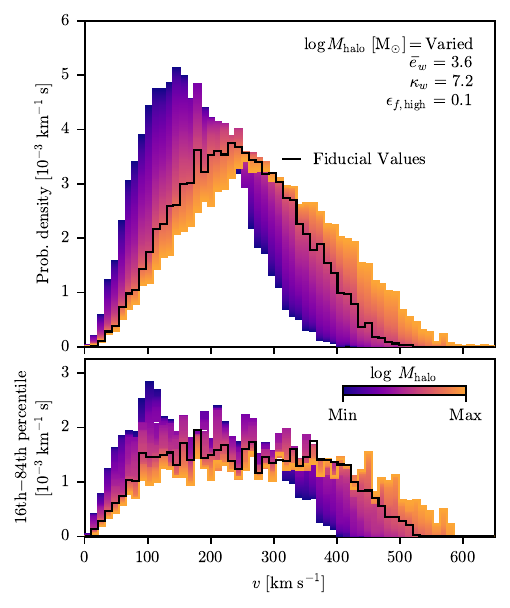}
  \caption{
Median DM speed distributions~(top panel) and 16th--84th percentile range~(bottom panel) for 400 different halo masses, fixing baryonic feedback parameters to their TNG fiducial values. The distributions are colored by halo mass from $\log\mhalo/\mathrm{M}_{\odot} = 11.6$~(Min; dark purple) to $12.2$~(Max; orange). The black line corresponds to the TNG fiducial values and $\log\mhalo/\mathrm{M}_{\odot} = 12$. As halo mass increases, the potential well deepens, and the median speed distribution widens and peaks at higher values. Similarly, the 16--84$^{\rm th}$ percentile range (at a given speed) grows at larger halo masses, especially above $\sim 400\kms$. }
  \label{fig:mhalo}
\end{figure}

Because the halo mass of a galaxy can strongly influence the shape of its potential well, it can also affect the local DM speed distribution. Using the emulator (see Section~\ref{sub:nf}), we generate 100 samples in 400 halo mass bins ranging from $\log\mhalo/\mathrm{M}_{\odot} = 11.6\text{--}12.2$ (for a total of $4\times 10^4$ samples), which encompasses the uncertainty for our own Galaxy~\citep{Bland_Hawthorn_2016}.\footnote{The choice of 100 samples per bin is intended to marginalize over the intrinsic halo-to-halo variance.  In testing, we confirm that increasing the number of samples beyond 100 does not impact the resultant speed distribution.} To isolate the effects of halo mass for this comparison, we fix the baryonic feedback parameters to their TNG fiducial values.

\begin{figure*}
\centering
\includegraphics{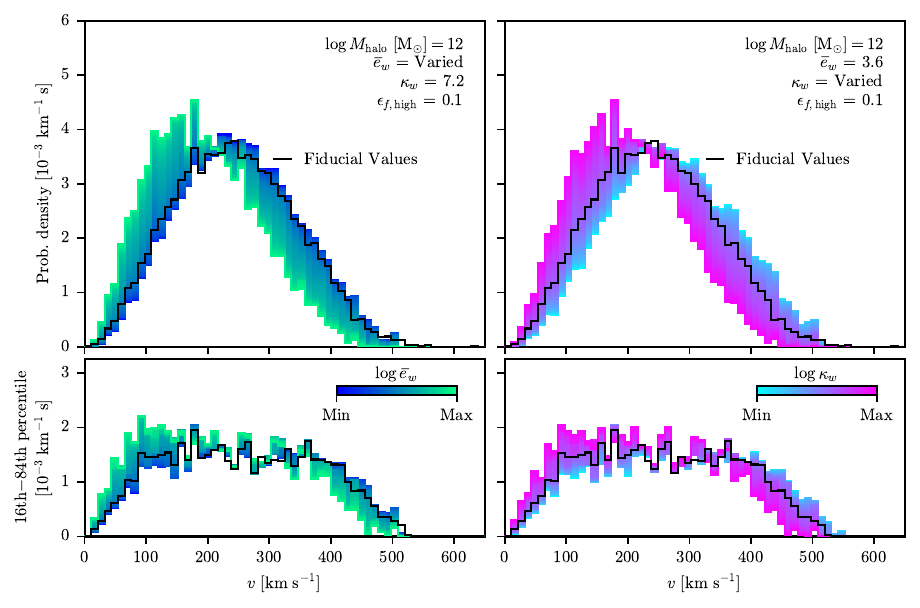}
  \caption{
 Median DM speed distributions~(top panel) and 16th--84th percentile range~(bottom panel) over 100 emulated samples each at 400 values of the supernovae parameters. In the left panel \ew is varied from 0.9~(dark blue) to 14.4~(light green). In the right panel $\kw$ is varied from 3.7~(light purple) to 14.8~(cyan). The black line shown uses the TNG fiducial values and $\log\mhalo/\mathrm{M}_{\odot} = 12$ and is the same as that in Figure~\ref{fig:mhalo}. As \ew increases, the speed distributions peak at lower values. The right panel shows a similar result for $\kw$. When these two supernova parameters increase in value, the galaxy's stellar mass decreases, which makes a shallower potential well, corresponding to lower DM speeds.}
  \label{fig:snfeedback}
\end{figure*}

Figure~\ref{fig:mhalo} shows the median speed distributions in the top panel, 
as well as the 16--84th percentile uncertainty at each speed in the bottom panel. The dark purple to orange corresponds to the transition from the minimum to the maximum range of $\log M_{\rm halo}$. Increasing the halo mass shifts the DM speed distribution to higher speeds, since the larger mass deepens the potential well. 
The peaks of the distributions, $v_{\rm LSR}$, range from $154\kms$ at the lowest masses to $266\kms$ at the highest masses. The values of $v_{99}$ range from $329\text{--}524\kms$. These two metrics demonstrate how widely the peak and the tail of the distribution can vary based on halo mass. 

Note the spread of the 16th--84th percentile range above 400 $\kms$, implying the high-speed tail of the speed distribution is very sensitive to the halo mass compared to the lower-end of the distribution. Quantifying how much it takes to transform the speed distribution with the smallest halo mass to that with the largest, we find the EMD between the two median distributions to be $117\kms$.

See Appendix~\ref{sec:appsup} for a similar analysis on the stellar mass. The dependence of the speed distribution on stellar mass is sub-dominant to halo mass. The result mirrors that for \ew and $\kw$, presented in the next subsection, as the changing supernovae parameters affect the stellar mass of a galaxy. 

\vspace{0.01in}
\subsection{Baryonic Feedback}
\label{sub:sn}

This subsection focuses on trends between the galaxy-formation model and DM speed distributions.
In general, \ew and $\kw$ (which scale the specific energy for supernovae winds and the supernovae wind speeds, respectively) have the strongest impact on the DM speed distributions, while \agn\!\!, $\Omega_{\rm m}$, and $\sigma_8$ do not significantly impact the results.
We thus primarily focus on the results from the supernova feedback variations here.

The median speed distributions as a function of \ew and $\kw$ are shown in the top panel of Figure~\ref{fig:snfeedback}. In each case, all other feedback parameters are set to their TNG fiducial values while $\log M_{\rm halo}/\Msun = 12$.  High/low values of \ew are shown in green/blue while high/low values of $\kw$ are shown in purple/cyan. The 16--84$^{\rm th}$ percentile range, at a given speed, is provided in the bottom panel. Overall, the variation  in  the speed distributions is much smaller compared to the case of varying halo mass.

The peak speed, $v_{\rm LSR}$, decreases from $241$ to $164\kms$~($261$ to $176\kms$) moving from the minimum to maximum of \ew($\kw$).
Additionally, the $99^{\rm th}$ percentile of the speed distribution, $v_{99}$, decreases from $460$ to $399\kms$~($491$ to $398\kms$) moving from the minimum to maximum of \ew($\kw$).
For the median distribution, the EMD between the distribution for the lowest \ew value and that for the highest is $58\kms$. For the scaling of wind speeds, $\kw$, the EMD between the  distribution for the lowest value and the highest is $74\kms$. Note how these values compare to the previous subsection discussing halo mass. The work required to transform a distribution from the low end of \ew to the high end is significantly less than the work to transform from the low end of $\mhalo$ to the high end, further reinforcing that the range of distributions that result from varying baryonic physics is much smaller compared to varying halo mass.
This further strengthens the conclusion that the uncertainty of the emulated Varied dataset in Figure~\ref{fig:milkyway} is dominated more by halo-to-halo variance and uncertainty in halo mass and not uncertainty in the baryonic parameters. 

As the supernova feedback parameters increase, the distributions shift towards lower speeds.  The underlying reason for this is that the stellar masses of galaxies decrease as \ew and $\kw$ increase---see~\cite{Garcia_to_Appear} for a more detailed discussion. Due to the non-linear nature of supernovae feedback, it is difficult to pinpoint the exact explanation for these trends. Possible explanations can be found in \cite{tng2}: increasing \ew increases the specific energy for supernovae winds, which increases the effective pressure for outflow of gas from the host potential and leads to less star-forming gas and lower stellar masses. Increasing $\kw$ increases the supernova wind speeds. Larger speeds enable gas to flow more easily and escape the host halo; gas recycling is reduced, and the available gas for star formation is lower.

As mentioned previously, we do not explicitly show variations with the other input parameters to the emulator. Briefly, varying the AGN feedback parameter, \agn\!\!\!, does not change the DM speed distribution. Specifically, we find EMD values of $4.9\kms$ between the median speed distribution for the lowest and highest \agn value. 
 For the cosmology parameters, the corresponding EMD values are $1.25\kms$ for $\Omega_{\rm m}$ and $3.34\kms$ for  $\sigma_{8}$.

\section{Discussion}
\label{sec:discus}

Throughout this section, we compare the DREAMS results against previous studies in the literature, in particular those using TNG50 and FIRE-2 galaxies. We then propagate these findings to direct detection constraints and discuss how different samples lead to different constraints on the scattering cross section.

\vspace{0.5in}
\subsection{Comparison to Previous Simulation Results}

This subsection contextualizes the results of this paper with previous literature. Specifically, 
we compare to~\cite{folsom2025darkmattervelocitydistributions}, which used the TNG50 simulation volume, and ~\cite{Staudt_2024}, which studied 12 FIRE-2 galaxies.
The former is an important validation of the results from Section~\ref{sec:results} as both DREAMS and TNG50 are built on the same subgrid model. The comparison with FIRE, on the other hand, tests the extent to which the detailed implementation of physics within the TNG model impacts the core results of this work.

\subsubsection{TNG50 Comparison}
\label{sub:tngcompare}

\begin{figure*}
\centering
\includegraphics{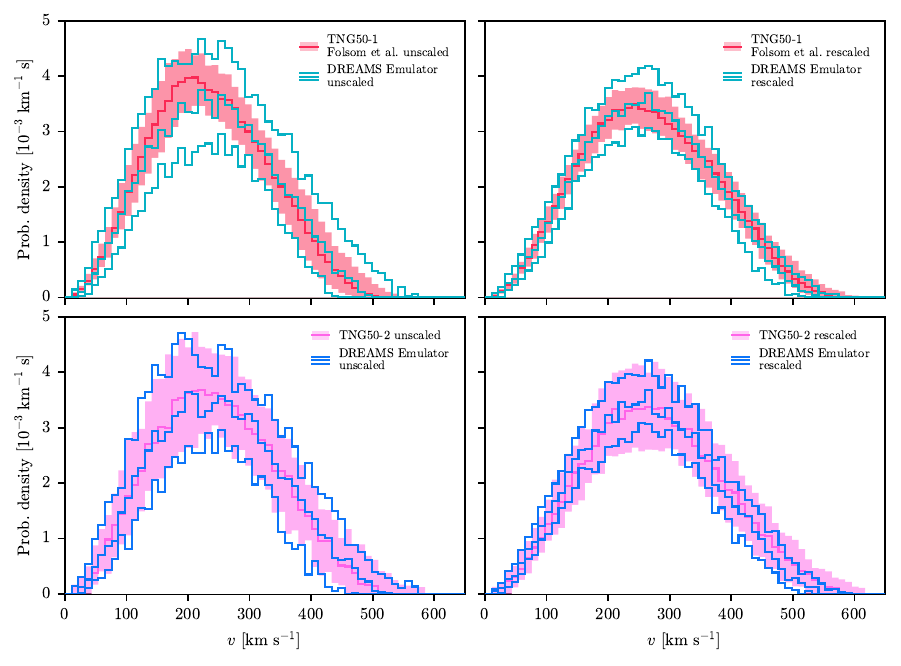}
  \caption{
DM speed distributions and 16th--84th ranges compared between DREAMS and TNG50. The top panels show \protect\cite{folsom2025darkmattervelocitydistributions} TNG50-1 results unscaled~(left panel) and rescaled~(right panel) alongside those of a new emulated DREAMS dataset where the number of galaxies, halo masses, stellar masses, and baryonic physics are matched exactly to the TNG50-1 galaxies (see discussion in Section~\ref{sub:tngcompare}). The bottom panels show the equivalent results for TNG50-2 galaxies, excluding any with $\log\mhalo/\Msunt > 12.2$ because DREAMS does not have halo masses this large and cannot emulate them reliably. For the rescaled galaxies, TNG50-1 has a smaller uncertainty band compared to DREAMS and TNG50-2 has a larger uncertainty band compared to DREAMS. The rescaling procedure rescales DM positions and velocities so the mass enclosed at the solar circle matches the mass enclosed for our MW.}
  \label{fig:tng50}
\end{figure*}

The TNG50-1 volume~\citep{nelson2021illustristngsimulationspublicdata,Nelson_2019,Pillepich_2019} employs the same sub-grid model as DREAMS and is at a similar resolution ($\sim \times2 \ \rm higher$).
\cite{folsom2025darkmattervelocitydistributions} selects 98 MW-mass galaxies from the volume by requiring them to have stellar mass between $(4\text{--}7.3)\times10^{10}\Msun$, be more than 500~kpc from any halo with larger mass, and be more than 1~Mpc from any halo with mass  greater than $10^{13}\Msun$. 
Their speed distributions are determined for DM in an annulus that extends 2~kpc in radius and 2~kpc in height, centered at a galactocentric radius of 8.3~kpc. However, this alone does not reproduce the correct peak speed of the MW.
To correct for this, \cite{folsom2025darkmattervelocitydistributions} employ an energy-conserving transformation to both positions and velocities to scale them based on the radius at which the enclosed mass matches that of our MW at 8.3~kpc. 

To make a meaningful comparison with \cite{folsom2025darkmattervelocitydistributions}, we must account for differences between the two sets of simulations. As shown in Figure~\ref{fig:mhalo}, the halo mass affects the speed distribution. The halo masses of the TNG50-1 galaxies used by \cite{folsom2025darkmattervelocitydistributions} fall in the range $\log \mhalo/\Msunt \in [11.8, 12.2] $, with the distribution peaking around 11.9.  
Meanwhile, the DREAMS distribution is uniform in the range of $\log \mhalo/\Msunt \in [11.6, 12.2]$.  To emulate a DREAMS dataset that is analogous to the Folsom TNG50 sample, we fix the baryonic physics to the fiducial TNG values (see Section~\ref{sub:dreams}).\footnote{Here, we use the KDE method on both sets of data. However, the results shown are not sensitive to doing this or the annulus approach. } We generate 98 samples with halo masses corresponding to the halo masses of the 98 TNG50 galaxies. For each of these realizations, we select galaxies that match the stellar mass of each TNG50 halo (within a tolerance of $\sim3\%$).  
We also repeat this procedure for the TNG50-2 volume, which has the same initial conditions as TNG50-1 but at $\sim8\times$ coarser resolution. This enables us to get a sense of resolution-based differences in the resulting speed distributions.

The top-left panel of Figure~\ref{fig:tng50} compares the TNG50-1 speed distributions with those of the emulated DREAMS sample. In this case, the TNG50-1 speeds are unscaled and the emulator is also trained on the unscaled DREAMS speeds (i.e., the standard simulation output). While the speed distributions are for the solar circle, there is no requirement that the simulated galaxies have a morphology that reproduces the $M_{\rm LSR}$ of the MW. This means that some of the galaxies may be too compact and others not enough. The DREAMS distributions peak at a higher speed than those in TNG50-1 and exhibit a wider spread. Specifically, for DREAMS we find $v_{\rm LSR}=232^{+47}_{-47}\kms$ and for the TNG50-1 galaxies, $215^{+44}_{-44}\kms$. These differences likely arise from the different galaxy morphologies between the two samples.

To better match galaxy morphologies, we use the procedure described in \cite{folsom2025darkmattervelocitydistributions}, which rescales galaxies to have the same enclosed mass at the solar radius as the MW (see \textit{Scaling phase space} and Appendix~A of \cite{folsom2025darkmattervelocitydistributions}). 
The right panel of Figure~\ref{fig:tng50} shows the rescaled TNG50-1 results.  To compare to these results, we generate a new emulated DREAMS dataset that is trained on the \emph{rescaled} velocities of the simulated galaxies (rather than the raw simulation outputs).  In every other way, this emulator is the same as what we used previously. The results of the rescaled dataset are also shown in the right panel of Figure~\ref{fig:tng50}.  
With the rescaling procedure, DREAMS and TNG50-1 agree very well, with the peaks of the distributions coming into alignment: $v_{\rm LSR} = 258^{+28}_{-28}\kms$ for DREAMS and $v_{\rm LSR} = 247^{+33}_{-33}\kms$ for TNG50-1. Note that the 16--84$^{\rm th}$ range for DREAMS tightens, but is still larger than that for TNG50-1, with a percent difference of about $20\%$ around the peak speeds. 

The bottom panel of Figure~\ref{fig:tng50} shows the corresponding results, except using TNG50-2. This uses the same isolation criteria as \cite{folsom2025darkmattervelocitydistributions} but on the TNG50-2 volume: this includes a total of 57 host galaxies.\footnote{We remove any TNG50-2 galaxies with $\log \mhalo/\Msunt > 12.2$ because the DREAMS training parameter space does not include this region and therefore would be unable to compare to these galaxies.} 
After the rescaling, the peaks of the TNG50-2 and emulated DREAMS dataset are in alignment; the 16--84$^{\rm th}$ range for DREAMS is however smaller than that for TNG50-2, with a percent difference of about $30\%$.

The comparisons in Figure~\ref{fig:tng50} allow us to account for the major differences in halo mass, stellar mass, baryonic physics, sample size, and morphologies between TNG50 and DREAMS in an even-handed way. The rescaled emulated DREAMS dataset has a smaller spread in distributions than TNG50-2, and TNG50-1 has a smaller spread than DREAMS. Recall that DREAMS has $\sim2\times$ lower resolution than TNG50-1 and TNG50-2 has $\sim8\times$ lower resolution than TNG50-1. Thus, the higher the resolution of a suite, the smaller the width of the spread in the DM speed distribution. While it is not clear what causes resolution to increase the spread, it is possible that it can affect the disk morphology as stellar mass is a resolution-dependent quantity \citep{Pillepich_2019}. However, this explanation is likely not the sole cause because we would expect a systematic shift in the distributions to higher/lower speeds. Another possible explanation is the numerical exchange of energy between DM and baryons (see \cite{Ludlow_1,Ludlow_2,Ludlow_3}) where DM takes angular momentum from gas particles. This may add velocity to DM particles and increase the spread in the distributions. In order to further probe these questions, an approach that varies over resolution is necessary. This would include either simulating the same galaxy multiple times at varying resolution, or a new DREAMS suite where resolution is a parameter that is varied over alongside baryonic physics.

The environment and formation history of the MWs in the DREAMS and TNG50 samples could also play a role. DREAMS has an order-of-magnitude more galaxies than TNG50-1, which are found in individual 145~Mpc boxes rather than one 50~Mpc box for TNG50-1. The isolation criteria of DREAMS (see Section~\ref{sub:DM_vd}) and \cite{folsom2025darkmattervelocitydistributions} are different.  For example, \cite{folsom2025darkmattervelocitydistributions} requires that galaxies are more than 500~kpc from any halo with larger mass and are more than 1~Mpc from any halo with mass  greater than $10^{13}\Msun$. DREAMS requires that there are no galaxies with mass greater than $1.0\times10^{12}\Msun$ within 1~Mpc. 

\subsubsection{FIRE Comparison}
\label{sub:firecompare}

\begin{figure}
  \centering
  \includegraphics{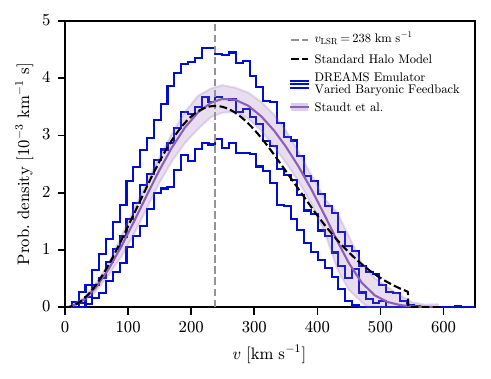}
  \caption{
Speed distribution for FIRE-2 results from \protect\cite{Staudt_2024} (including their one standard deviation uncertainty band, which includes uncertainty propagated from the MW circular velocity and the root mean square of the difference between their model and their simulation data) compared with DREAMS results when matching the circular velocities between the two suites. The dark blue emulated DREAMS curve includes $10^{5}$ samples and shows the median speed distribution and 16th--84th band. The FIRE-2 result is contained within the uncertainty of the DREAMS result and peaks at higher speeds, hinting at some difference due to e.g., resolution, sub-grid physics, or sample size. The DREAMS dataset corresponds to $\log \mhalo/\Msunt \in [12,12.2]$ and $ 10.60 \leq \log\mstar/\Msunt \leq 10.78$ and $  10.972 \leq \log\mlsr/\Msunt \leq  11.027$. The SHM is shown in dashed black and $v_{\rm LSR}$ is denoted by the gray-dashed line. The FIRE-2 results falls within the DREAMS band, but the spread of the latter is much larger. }
  \label{fig:fire}
\end{figure}

As with the comparison to TNG50, we now compare our results to \cite{Staudt_2024} in a systematic manner. This comparison tests the impact of different feedback models (FIRE versus TNG). \cite{Staudt_2024} uses 12 FIRE-2~\citep{Hopkins_2018} galaxies to infer the local speed distribution with parameters that depend on the local circular speed of the galaxy. They use a MW circular velocity of $229\pm7\kms$ and define circular velocity as the average azimuthal velocity of gas particles with temperatures below $10^{4}$~K. As shown by \cite{Staudt_2024}, this has a one-to-one relationship with the  spherical approximation for the circular velocity, defined as
\begin{equation} v_{\mathrm{circ}} = \sqrt{\frac{G M(R)}{R}} \, ,
\label{eq:vcirc}
\end{equation}
where $G$ is Newton's constant, $M(R)$ is the mass enclosed at $R$, and $R$ is the radius from the center of the galaxy.  To select a similar subset of galaxies within DREAMS, we use Equation~\ref{eq:vcirc}, since there are no gas particles with temperatures below $10^{4}$~K.\footnote{In the DREAMS suite, the average stellar speeds as a function of mass enclosed coincide strongly with a spread around the spherical approximation with $\sim68\%$ of galaxies having a difference between average stellar speed and the spherical approximation of less than 13.85$\kms$ and $\sim95\%$ of galaxies having a difference of less than 40.15$\kms$. This justifies our use of the spherical approximation (Equation~\ref{eq:vcirc}).}
 Specifically, we convert the circular velocity used by \cite{Staudt_2024} to a range of $\mlsr$ to compare to DREAMS. This corresponds to $\log\mlsr/\Msunt \in [10.972, 11.027]$. 

In detail, we use the emulator trained for the DREAMS suite with Varied baryonic physics, introduced in Section~\ref{sec:results}.  Note that this is \emph{not} the same as the emulator used in Section~\ref{sub:tngcompare}, which is specific to TNG50.  We generate $10^5$ realizations, restricting the halo masses to the range $\log \mhalo/\Msunt \in [12,12.2]$ (consistent with the range of masses of the \cite{Staudt_2024} sample) and $\log\mlsr/\Msunt \in [10.972, 11.027]$, and varying over the full range of DREAMS baryonic feedback parameters. 

Figure~\ref{fig:fire} shows the emulated DREAMS result in dark blue and the FIRE-2 result in purple. The speed distributions for the emulated DREAMS dataset peak at lower values compared to those of \cite{Staudt_2024}. Their 12 galaxies have DM speed distributions that are peaked to the right of the stellar circular velocities, but this is not the case for the DREAMS suite.
The 16--84$^{\rm th}$ range is also notably larger for the DREAMS result, encompassing the entirety of the \cite{Staudt_2024} result. There are several possible explanations for these discrepancies.

The first is the difference in the number of galaxies included in the band: \cite{Staudt_2024} studies 12 galaxies while the emulated DREAMS dataset includes $10^{6}$. Subselecting the emulated DREAMS dataset by 12 galaxies and plotting the distribution, the results typically fall within the dark blue band.  Indeed, if we take many realizations of 12 subsamples and plot the resulting 16-50-84$^{\rm th}$ percentiles, we essentially recover the dark blue band in Figure~\ref{fig:fire}.  This strongly suggests that the spread of the FIRE distribution undersamples halo-to-halo variance.  By construction, the DREAMS dataset does sample over the initial conditions of a much larger number of MW-mass systems, and the overall uncertainty on the speed distribution increases.

That being said, there are several other factors that can also contribute to the difference in uncertainty and it is not possible to fully disentangle these effects at this stage.  For example, 
there is a resolution difference between the two suites.  Six of the FIRE-2 galaxies have a DM particle mass resolution of $3.5 \times 10^{4}\Msun$ and six have $2 \times 10^{4}\Msun$ (almost the difference in resolution between DREAMS and TNG50, but at two orders-of-magnitude higher resolution than DREAMS). As the FIRE-2 galaxies have a lower resolution, and given the discussion of Section~\ref{sub:tngcompare}, it is possible that the FIRE-2 result has a tighter uncertainty band as a result of using higher- resolution simulations. Further analysis of resolution effects, using a suite where resolution is varied for example, is necessary to understand the difference that resolution makes here.

Additionally, the different sub-grid model can potentially be contributing to the differences. This can be tested in future work by generating a new MW-mass zoom-in suite for DREAMS that uses the FIRE-2 model. Such a suite, which would be generated for a large range of halo initial conditions, would also help to disentangle the effects of different MW environments.  

\vspace{0.3in}
\subsection{Applications to Direct Detection}
\label{sub:directdetection}
The local DM speed distribution feeds into the scattering rates predicted for direct detection experiments.  This subsection explores this dependence in further detail for the DREAMS predictions, focusing on the example of elastic DM-nucleon scattering. We compute a mock recoil spectrum for each DREAMS halo using the \texttt{wimprates}~\citep{wimprates} package. Following \cite{folsom2025darkmattervelocitydistributions}, the recoil spectrum is passed through an approximate likelihood model for the XENON1T analysis~\citep{xenon1t} which constrains the spin-independent DM-nucleon scattering cross section, $\sigma_{\rm SI}$, to the value that would exceed the modeled detector background at 90\% confidence level~(CL). 

\begin{figure}
\centering
\includegraphics{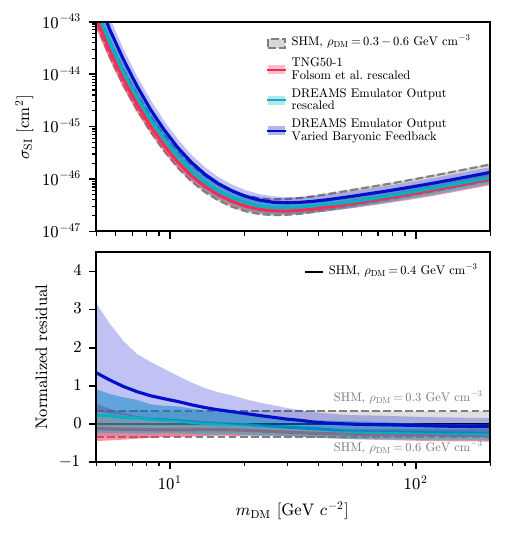}
  \caption{
 The top panel shows DM–nucleon spin-independent cross-section limits as a function of fundamental DM mass, $m_{\rm DM}$, for the XENON1T experiment. The limits are provided for the TNG50-1 result \protect{\citep[see also][]{folsom2025darkmattervelocitydistributions}} shown in Figure~\ref{fig:tng50} (red), the emulated rescaled DREAMS result shown in Figure~\ref{fig:tng50}~(cyan), and the emulated DREAMS dataset with Varied baryonic physics shown in Figure~\ref{fig:milkyway}~(dark blue). The curves correspond to the constraint at 90\% confidence with 16--84$^{\rm th}$ uncertainty bands. We also show the SHM result with $\rho_{\rm DM}$ varying from $0.3\text{--}0.6$~GeV~cm$^{-3}$. The bottom panel shows the normalized residual to the SHM at $\rho_{\rm DM} = 0.4 \mathrm{\ GeV \ cm}^{-3}$~(black). Comparing the red and cyan bands underscores that the simulation resolution can affect sensitivity projections. From low $m_{\rm DM}$ to high $m_{\rm DM}$, the uncertainty band widths vary from $\sim 0.23\text{--}0.08$~dex, $\sim 0.19\text{--}0.06$~dex, and $\sim 0.15\text{--}0.12$~dex for DREAMS Varied, TNG50-1 rescaled, and DREAMS rescaled, respectively. At low masses, $m_{\rm DM} \lesssim 10 \mathrm{\ GeV} \ c^{-2}$, the limit for the DREAMS Varied baryonic physics result is weaker compared to the other two datasets as the high-speed tail is suppressed, due to  baryonic physics variations, simulation resolution, halo-to-halo variance, and halo mass uncertainty. The DREAMS result falls within the $2\sigma$ XENON1T sensitivity band in \cite{xenon_exp}.  } 
  \label{fig:dd}
\end{figure}

There are two key updates to the emulator used in this subsection. First, we train it to predict the binned heliocentric speed distribution, as this is the input taken by \texttt{wimprates}. In particular, the emulator predicts the speed in 90 bins, from $0\text{--}975\kms$, using the same bin width as the distribution for the galactic frame.
Second, the emulator also outputs the local DM density, $\rho_{\rm DM}$, which also affects the scattering rate. Appendix~\ref{sec:appemul} provides validation for this emulator; see \cite{Garcia_to_Appear} for a more in-depth discussion of the DM density profiles of the DREAMS MW-mass halos.

The top panel of Figure~\ref{fig:dd} shows the cross-section limit for DM–nucleon scattering for the XENON1T experiment, as
a function of the fundamental DM mass, $m_{\rm DM}$. These curves show the direct detection constraint at 90\% confidence with 16--84$^{\rm th}$ uncertainty bands for three different sets of samples. The limits are shown for the emulated DREAMS data with Varied baryonic physics in dark blue (corresponding to Figure~\ref{fig:milkyway}), as well as the rescaled TNG50-1 result and emulated DREAMS TNG50-1 result in red and cyan, respectively (corresponding to  Figure~\ref{fig:tng50}). The gray band corresponds to the direct detection constraint assuming the SHM and varying over the experimental uncertainty of $\rho_{\rm DM}$. The bottom panel of Figure~\ref{fig:dd} shows the limit from each dataset normalized to the SHM with the local DM density set to $\rho_{\rm DM} = 0.4 \mathrm{\ GeV \ cm}^{-3}$. 

For DM masses greater than 25~GeV~$c^{-2}$, the dominant uncertainty on the scattering rate comes from $\rho_{\rm DM}$. Therefore, the differences between the limits can be explained by the differences in the local DM density. The emulated DREAMS dataset with Varied baryonic feedback has $\rho_{\rm DM} = 0.38^{+0.099}_{-0.071}\mathrm{\ GeV \ cm}^{-3}$, which falls within observational uncertainties for the MW~\citep{density_salas}.  For comparison, the rescaled TNG50-1 galaxies have $\rho_{\rm DM} = 0.51^{+0.13}_{-0.07} \mathrm{\ GeV \ cm}^{-3}$ and the emulated rescaled DREAMS dataset has  $\rho_{\rm DM} = 0.49^{+0.13}_{-0.12}\mathrm{\ GeV \ cm}^{-3}$. The limits become slightly ($\lesssim 10\%$ at high $m_{\rm DM}$) weaker moving from the TNG50-1 set to the emulated DREAMS Varied set because the $\rho_{\rm DM}$ values become correspondingly smaller. 

At the low DM masses, $m_{\rm DM} \lesssim 10 \mathrm{\ GeV} \ c^{-2}$, only high-speed DM particles are detected and thus the main uncertainty on the limit comes from the tail of the DM speed distribution. Comparing the rescaled TNG50-1 and emulated DREAMS dataset (red and cyan bands, respectively), the latter is shifted towards higher values of $\sigma_{\rm SI}$ and with a larger uncertainty. This follows from the speed distribution in the upper-right panel of Figure~\ref{fig:tng50}, where the 16$^{\rm th}$ percentile of the DREAMS rescaled result falls faster than that of TNG50-1 towards high speeds. Recall that the main difference between these two datasets is the resolution of the simulations on which they are based.  This therefore demonstrates how resolution effects in speed distributions derived from simulations can propagate to sensitivity projections for direct detection experiments. 

The emulated DREAMS dataset for Varied baryonic physics (blue band) yields the weakest limit at low $m_{\rm DM}$. This leads to a $\lesssim 20\%$ weaker constraint at low masses relative to TNG50-1. This is because the varied baryonic feedback result has a tail that is suppressed compared to the other two results, as seen in Figure~\ref{fig:milkyway}. Specifically, the Varied DREAMS result has $v_{99} = 498^{+32}_{-32} \kms$, compared to the value for the TNG50-1 rescaled, $v_{99} = 505^{+32}_{-24} \kms$. This difference is likely due to a combination of halo mass uncertainty, halo-to-halo variance, varying baryonic physics, and resolution compared to the TNG50-1 speed distributions.

The three datasets considered here lead to sensitivity projects that are all comparable to the projected sensitivity bands for XENON1T \citep{xenon_exp}. Specifically, the TNG50-1 and emulated DREAMS rescaled datasets are consistent within $1\sigma$ and the emulated DREAMS Varied dataset is consistent within $2\sigma$. This reinforces the conclusion of \cite{folsom2025darkmattervelocitydistributions}, which set the astrophysical uncertainties on direct detection analyses on the same footing as the experimental ones for the first time.

\section{Conclusions}
\label{sec:concl}
This paper used the DREAMS CDM suite of 1024 MW-mass zoom-in simulations to analyze the DM speed distribution near the solar position.  We utilized a novel KDE method to recover the distribution, and implemented machine-learning approaches to disentangle the impact of halo-to-halo variance, halo-mass uncertainty, and baryonic feedback modeling on the final result.  We obtained a state-of-the-art prediction for the local DM speed distribution for the MW, comparing it to prior studies in the literature and studying its impact on limits for the XENON1T direct detection experiment. 

Specifically, we presented a speed distribution and uncertainty band for simulated MW-mass halos (Figure~\ref{fig:milkyway}) accounting for uncertainties in the halo mass (Figure~\ref{fig:mhalo}) and galaxy-formation model (Figure~\ref{fig:snfeedback}). The distribution restricts to galaxies with stellar masses consistent with that of the MW. To account for the compactness of the MW, we also restricted the mass enclosed at the solar radius, 8.3 kpc, to within the uncertainty on the MW value. Increasing halo mass shifts the speed distributions to higher values, as expected because of the deeper potential.  The effects are considerably weaker when varying the baryonic feedback parameters within the TNG model.  Increasing supernovae feedback tends to decrease the stellar mass of galaxies, which shifts the speed distributions to lower values.  However, the magnitude of these shifts are smaller than what is observed with shifts in host halo mass. Black hole feedback, as well as variations on cosmological parameters, have no significant impact. 

Across the four parameters $\mhalo$, \ew\!\!, $\kw$, and $\epsilon_{f,\mathrm{high}}$, the uncertainty in our MW result is dominated primarily by halo-to-halo variance and uncertainty on the value of $\mhalo$ over contributions from feedback modeling. Overall, halo-to-halo variance is the primary contributor to uncertainty on the speed distribution, with halo mass contributing only $\mathcal{O}(10\%)$ to the overall scatter in quantities like $v_{\rm peak}$ and $v_{99}$~(Figure~\ref{fig:emul_val_2}). This suggests that observational improvements on the MW halo mass can help decrease, but will not ameliorate, the uncertainty on the local DM speed distribution.

We compared to two other TNG-based sets of simulations, TNG50-1 and TNG50-2, and the results of \cite{folsom2025darkmattervelocitydistributions}.  Once differences between baryonic physics, sample size, halo mass, stellar mass, and galaxy morphology were accounted for, we found good agreement with the DREAMS results.  This comparison demonstrated how particle mass  resolution can impact the DM speed distributions recovered from simulations, with lower resolutions inducing larger scatter in the predictions~(Figure~\ref{fig:tng50}). We also compared our results to the work of \cite{Staudt_2024}, which analyzed 12 FIRE-2 galaxies. The DREAMS results are consistent with the FIRE results, albeit with significantly larger scatter because they better sample the halo-to-halo variance by including nearly $100\times$ more galaxies.  Other effects such as resolution and choice of sub-grid physics may also play a role, but are not possible to disentangle at this stage. 

To illustrate the applicability of these results to direct detection experiments,  we demonstrated that the DREAMS speed distribution can be used to obtain projected limits for the XENON1T experiment~(Figure~\ref{fig:dd}). The uncertainty on the limit, which is dominated by halo-to-halo variance, falls within two standard deviations of the XENON1T sensitivity band~\citep{xenon1t}.  This is larger, but still comparable, to the uncertainty predicted using the TNG50-1 speed distribution~\citep{folsom2025darkmattervelocitydistributions}.  The difference is likely due to the fact that the DREAMS simulations are run at lower resolution, marginalize over baryonic feedback uncertainty, have different halo-mass distributions, and better sample halo-to-halo variance. 

There are three concrete extensions to the current DREAMS framework that can be undertaken to further solidify the predictions for the local DM speed distribution in the MW.  The first is to better quantify the effects of simulation resolution on these predictions. This can be achieved by generating a new suite of MW-mass zoom ins where resolution is included as a varied parameter, similar to the cosmological and astrophysical  parameters currently considered.  An emulator can then be used to learn the connection between resolution and the local DM speed distribution.

The second extension regards restricting the simulation suite to MW-mass galaxies with e.g., LMC analogs or GSE mergers to study whether these known events bias the predictions in certain directions.  (See  \cite{Rose_Centrals_to_Appear_A} for a discussion of GSE mergers in DREAMS.) Similarly, due to the isolation criteria of the current DREAMS suite, the MW-mass galaxies do not have large neighbors like Andromeda; future work could lift these requirements to allow for massive neighbors.

The third extension is to vary the baryonic feedback prescription.  The current DREAMS suite uses the TNG feedback physics, but models such as FIRE~\citep{Hopkins_2018} are an interesting counterpoint, as they produce bursty feedback which is not seen in TNG. Repeating the analysis on these simulations could further elucidate the uncertainty between feedback models and how it compares to the uncertainty from varying individual model parameters. 

The speed distributions procured in this work and the code used to produce them are publicly available. These results can be utilized in sensitivity projections or experimental analyses to robustly propagate theoretical uncertainties on the local DM phase space into direct detection results.

\vspace{2.5in}
\section*{Acknowledgments}
The authors gratefully acknowledge the use of computational resources and support provided by the Scientific Computing Core at the Flatiron Institute, a division of the Simons Foundation. They also acknowledge Research Computing at The University of Virginia for providing computational resources and technical support that have contributed to the results reported within this publication (URL: \url{https://rc.virginia.edu}), as well as Princeton University's Research Computing resources.

EL acknowledges many helpful discussions with Dylan Folsom as well as shared code and data which helped the production of Figure~\ref{fig:tng50} and Figure~\ref{fig:dd}. ML is supported by the Simons Investigator Award. AMG, AF, NK, and PT acknowledge support from the National Science Foundation under Cooperative Agreement 2421782 and the Simons Foundation grant MPS-AI-00010515 awarded to NSF-Simons AI Institute for Cosmic Origins — CosmicAI, \href{https://www.cosmicai.org/}{https://www.cosmicai.org/}. KEK is supported by the National Science Foundation Graduate Research Fellowship Program under Grant No.~DGE-2444107. 
XS acknowledges the support of the NASA theory grant JWST-AR-04814.  We thank the Simons Foundation for their support in hosting and organizing workshops on the DREAMS Project. 

{\em Software}: 
\texttt{Matplotlib}~\citep{Hunter2007},
\texttt{NumPy}~\citep{vanwderwalt2011},
\texttt{Pytorch}~\citep{pytorch},
\texttt{PyTorch Lightning}~\citep{lightning},
\texttt{Scipy}~\citep{scipy},
\texttt{wimprates}~\citep{wimprates},
\texttt{zuko}~\citep{zuko}.

\section*{Data Availability}
The code used to calculate KDE speed distributions and emulated speed distributions for this paper is available at \url{https://github.com/ejlilie/DREAMS_velocity_distributions}. Additional data will be shared on reasonable request to the corresponding author.

\clearpage

\appendix

\section{Kernel Density Estimation}
\label{sec:appkde}

\setcounter{equation}{0}
\setcounter{figure}{0} 
\setcounter{table}{0}
\renewcommand{\theequation}{A\arabic{equation}}
\renewcommand{\thefigure}{A\arabic{figure}}
\renewcommand{\thetable}{A\arabic{table}}

In this work, we use Kernel Density Estimation~(KDE) to determine the dark matter~(DM) speed distribution for a simulated Milky Way~(MW)-mass halo. This method weighs simulation particles based on their distance to the solar circle. As described in  Section~\ref{sub:DM_vd}, \texttt{Scipy.spatial.KDTree}~\citep{scipy} is used to calculate any number of nearest particles to a sample point and their distance from that point. These distances are then put into the kernel function $W(r,h)$, defined in Equation~\ref{eq:kernel}, and the output of the kernel function is the weight of that particle in the speed histogram. This is done for many points around the solar circle and then summed to produce the histogram of total speeds. 

\begin{figure*}[b]
\centering
\includegraphics{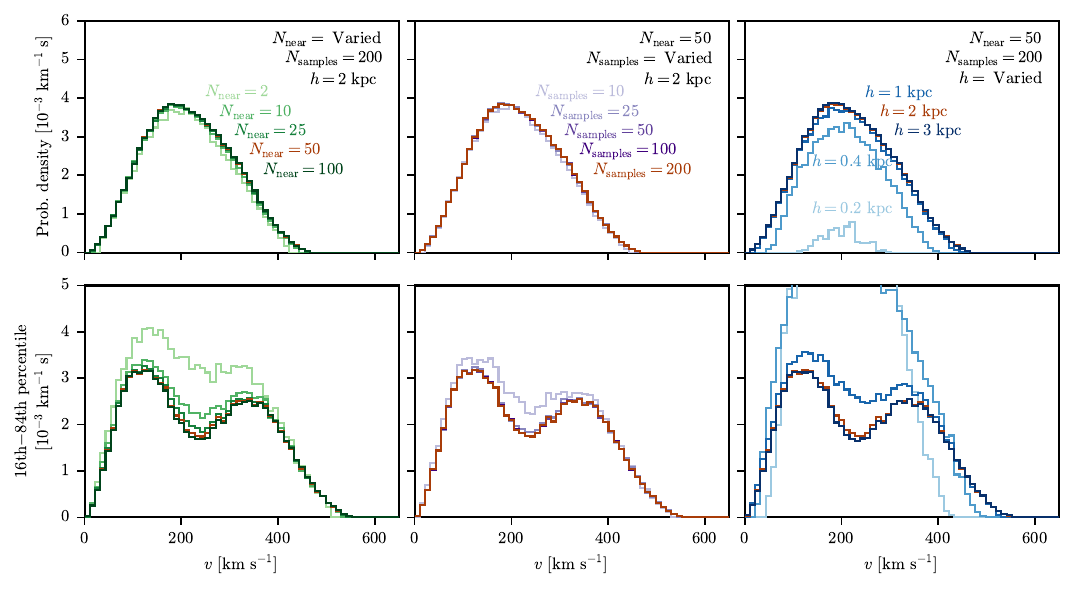}
  \caption{
Median DREAMS simulation KDE~(top row) and 16th--84th uncertainty range~(bottom row) for different values of the KDE parameters. From left-to-right, the columns correspond to variations in  $N_{\mathrm{near}}$, $N_{\mathrm{samples}}$, and $h$, respectively (holding the other parameters fixed). Across all three columns, the median distributions converge before the width of the uncertainty does. The fiducial values used in this work correspond to the brown curve in each panel with $N_{\mathrm{near}} = 50$, $N_{\mathrm{samples}} = 200$, and $h$ = 2 kpc.  }
  \label{fig:kde}
\end{figure*}

The KDE parameters used in this paper are the number of sample points along the solar circle, $N_{\mathrm{samples}}$, the number of nearby particles used for each sample point, $N_{\mathrm{near}}$, and the width of the kernel function, $h$. Figure~\ref{fig:kde} provides the median speed distributions~(top row) and 16--84th ranges~(bottom row) across the 1024 DREAMS simulations for different values of $N_{\rm samples}$~(left panels), $N_{\rm near}$~(middle panels), and $h$~(right panels). The left panel shows the speed distributions and 16th--8th percentile uncertainty bands for different values of $N_{\rm samples}$, whereas the middle and right panels show the same except varying over different values of $N_{\rm samples}$ and $h$, respectively.  Values of $h$ outside of $\sim 3\times$ the softening length of the simulations lead to equivalent results. Similarly, values of $N_{\mathrm{near}}$ and $N_{\mathrm{samples}}$ above 2 and 25 particles, respectively, lead to equivalent results. The values we adopt in this paper are $h = 2 \rm \ kpc$, $N_{\mathrm{near}} = 50$, and $N_{\mathrm{samples}} = 200$ (highlighted in brown in the figure panels).

To ensure that the fiducial values are reasonable and minimize generalization error, we cross-validate the results using the Kolmogorov–Smirnov test. For any individual galaxy, we split the points along the solar circle randomly (referred to as sample points in Section~\ref{sub:DM_vd}) into training~(80\%) and validation~(20\%). We calculate two speed distributions for the training points and the validation points and then compute the Kolmogorov–Smirnov statistic, $D$: 
\begin{equation}
D = \sup_{v}|F_{\rm train}(v) - F_{\rm val}(v)| \ ,
\end{equation}
where $F_{\rm train}(v)$ is the cumulative distribution function for the training speed distribution and $F_{\rm val}(v)$ is the cumulative distribution function for the validation speed distribution. We then calculate this for various values of $h$, $N_{\rm near}$, and $N_{\rm samples}$ and look where $D$ is minimized. We further repeat this through 10-fold cross-validation; this includes repeating the calculation of $D$ 10 times for a random set of training and validation points, and then taking the median and 16th--84th percentile to get an uncertainty on the the values of $h$, $N_{\rm near}$, and $N_{\rm samples}$. For example, for the first DREAMS galaxy, $h = 2.34^{+1.4}_{-0.9} \ \rm kpc$, $N_{\rm near} = 49_{-35}^{+33}$, and $N_{\rm samples} = 367^{+86}_{-183}$. All DREAMS galaxies are consistent with the fiducial values of $h = 2 \rm \ kpc$, $N_{\mathrm{near}} = 50$, and $N_{\mathrm{samples}} = 200$.

\section{NeHOD Emulator Validation}
\label{sec:appemul}

\setcounter{equation}{0}
\setcounter{figure}{0} 
\setcounter{table}{0}
\renewcommand{\theequation}{B\arabic{equation}}
\renewcommand{\thefigure}{B\arabic{figure}}
\renewcommand{\thetable}{B\arabic{table}}

The NeHOD emulator~\citep{nguyen2024dreamsmadeemulatingsatellite} includes a Normalizing Flows model and a Variational Diffusion Model.  This work utilizes the Normalizing Flows model, described in Section~\ref{sub:nf}, which can generate speed distributions and MW properties for any set of \ew\!\!, $\kw$, \agn\!\!, and $\mhalo$. The emulator uses \texttt{PyTorch}~\citep{pytorch} to convert the dataset to \texttt{PyTorch} tensor objects. \texttt{zuko}~\citep{zuko} is used to define the Normalizing Flows Model and \texttt{PyTorch Lightning}~\citep{lightning} trains and validates the model and contains the AdamW gradient descent optimizer~\citep{adamw1,adamw2}. A weight decay coefficient of 0.01 and a peak learning rate of 0.0005 is used for the AdamW optimizer. For these two parameters, no hyperparameter tuning is done.

The galaxies are partitioned between (1)~training (80\% of the galaxies), where the emulator is learning the parameters for the model; (2)~validation (10\% of the galaxies), where the emulator performance is assessed and tuned; and (3)~testing (10\% of the galaxies), where the emulator is compared to a subset of the data it has not seen before. Training is done over 250 epochs with a maximum of 5000 training steps. Testing is not used to update parameters, but to see how the emulator performs on unseen data. The hyperparameters are the number of transforms (4), the number of projection dimensions (32 per transform), the number of hidden dimensions (16 per transform), and the dropout rate (0.1). When these values are varied in the ranges [4, 8], [32, 64],[16, 32], and [0.05, 0.2], respectively, the conclusions of our analysis are unchanged.

\begin{figure*}
  \centering
  \includegraphics{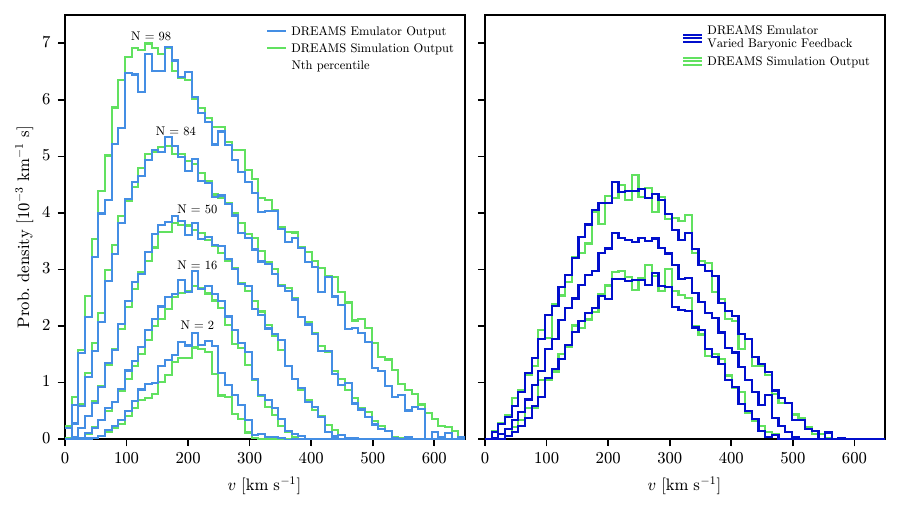}
  \caption{\emph{(Left)} DM speed distributions for the raw simulation outputs in green compared to the emulator output in blue when matching the exact DREAMS halo masses and feedback parameters for each simulated galaxy. The distribution is shown for the $N=[2,16,50,84,98]$ percentiles. The emulator reproduces the simulation results with the most disagreement at the peak and tail of the lowest/highest percentile. \emph{(Right)} The 16-50-84th percentiles for the raw simulation outputs in green, compared to the emulated DREAMS Varied dataset from Figure~\ref{fig:milkyway} in dark blue.  Note that the simulated galaxies are required to satisfy $ 10.60 \leq \log\mstar/\Msunt \leq 10.78$ and $ 10.975 \leq \log\mlsr/\Msunt \leq  11.085$ (53 galaxies).  There is excellent agreement between the two.}
  \label{fig:emul}
\end{figure*}

The benchmark emulator used in this paper inputs values of \ew\!\!, $\kw$, \agn\!\!, and $\mhalo$ and outputs 60 histogram bins for speeds from $0\kms $ to $650\kms$. The emulator does not inherently normalize the histogram results, but they can be normalized after they are output. To validate the results, we generate an emulated dataset that matches the $\mhalo$ and feedback parameters for each galaxy in the DREAMS suite.  The left panel of Figure~\ref{fig:emul} shows the resulting speed distribution for the DREAMS simulation suite in green, marked by the $N=[2,16,50,84, 98]$ percentiles. The blue lines show the corresponding percentiles for the emulated sample of galaxies.  The distributions agree well with each other, except at the $N=2$ and $N=98$ percentile level, where there are visible deviations in the peak and tails of the distribution.  As a separate validation test, we also compare the speed distribution of the emulated DREAMS Varied dataset (from Figure~\ref{fig:milkyway}) to the direct simulation output. The right panel of Figure~\ref{fig:emul} shows their agreement.  

\begin{figure*}
  \centering
  \includegraphics{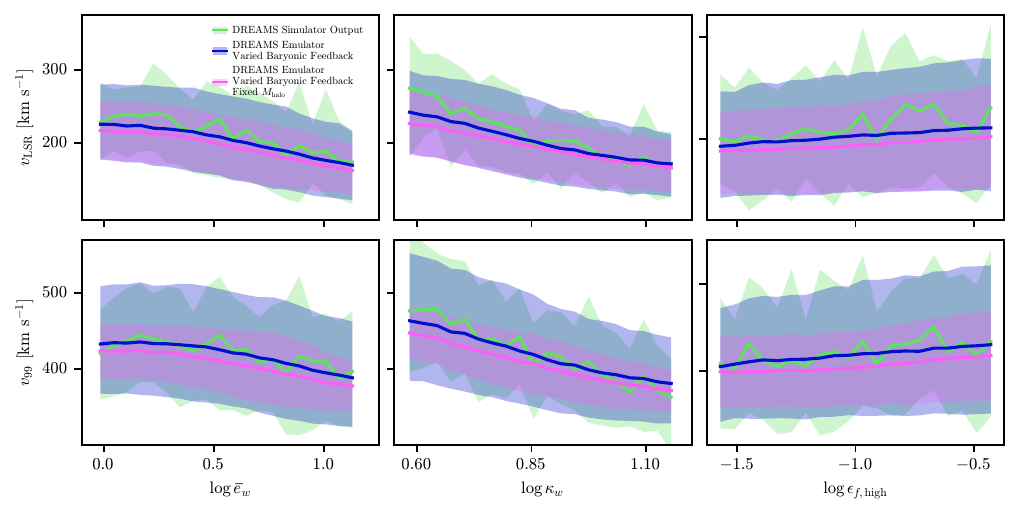}
  \caption{
Peak speed $v_{\rm LSR}$ (top panels) and 99th percentile speed $v_{99}$ (bottom panels) as functions of \ew (left), $\kw$ (middle), and \agn (right). The median and the 16th--84th uncertainty band for the simulation output is shown in green and the DREAMS Varied dataset (which marginalizes over halo masses) is shown in dark blue. In pink, we also show the DREAMS Varied dataset except restricted to $\log\mhalo/\Msunt$ to 11.9. This value is chosen as it is the midpoint of the DREAMS mass range. The emulator is able to learn the proper relationship between the speed distribution parameters and the baryonic parameters. The dominate source of uncertainty is halo-to-halo variance, however the uncertainty in halo mass can contribute up to a $\sim 30\%$ difference.}
  \label{fig:emul_val_2}
\end{figure*}

Figure~\ref{fig:emul_val_2} demonstrates how the peak speed, $v_{\rm LSR}$, and the 99th-percentile speed, $v_{99}$, depend on \ew, $\kw$, and \agn (while marginalizing over all other parameters) for both raw simulation output~(green) and the emulated DREAMS Varied dataset~(dark blue).  The emulated results adequately follow the median and 16--84$^{\rm th}$ percentile spread seen in the data, as desired.  Additionally, we show how the emulated DREAMS Varied dataset changes when fixing $\log \mhalo/\Msun $ to $11.9$~(pink). When fixing halo mass, the uncertainty band tightens. In particular, while the spread in $v_{\rm LSR}$ and $v_{99}$ is largely dominated by halo-to-halo variance, there is a subdominant contribution from uncertainty in halo mass that can induce as large as a $\sim 30\%$ difference. 

\begin{figure*}
  \centering
  \includegraphics{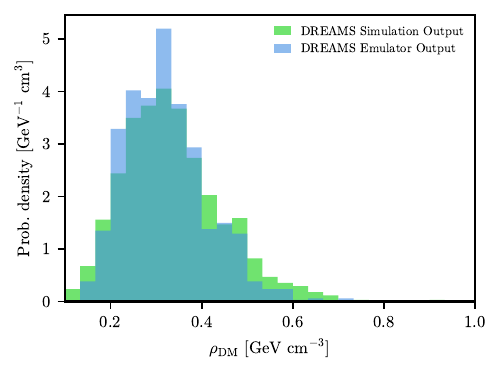}
  \caption{ Comparison of the local DM densities, $\rho_{\rm DM}$, for the raw simulation output~(green) and an emulated dataset that is matched to the $M_{\rm halo}$ and feedback parameters of the original 1024 galaxies~(blue). The EMD between the two distributions is $0.010\mathrm{\ GeV \ cm}^{-3}$.}
  \label{fig:densities}
\end{figure*}

Section~\ref{sub:directdetection} uses the emulated DREAMS speed distribution to present results for direction detection experiments.  For this calculation, we also emulated the local DM density, $\rho_{\rm DM}$, which is needed to determine experimental scattering rates. Figure~\ref{fig:densities} compares the distribution of $\rho_{\rm DM}$ taken directly from the DREAMS simulations with the results emulating the exact $M_{\rm halo}$ and baryonic parameters as the 1024 galaxies.  The EMD between the actual distribution and the emulated distribution is 0.010 GeV cm$^{-3}$. The emulator is clearly able to reconstruct the simulation results except in the extreme values of $\rho_{\rm DM}$.

\clearpage
\section{Supplementary Figures}
\label{sec:appsup}

\setcounter{equation}{0}
\setcounter{figure}{0} 
\setcounter{table}{0}
\renewcommand{\theequation}{C\arabic{equation}}
\renewcommand{\thefigure}{C\arabic{figure}}
\renewcommand{\thetable}{C\arabic{table}}

This appendix provides supplementary figures that complement the discussion in the main text. 
\vspace{0.2in}
\begin{figure*}[h]
\centering
\includegraphics{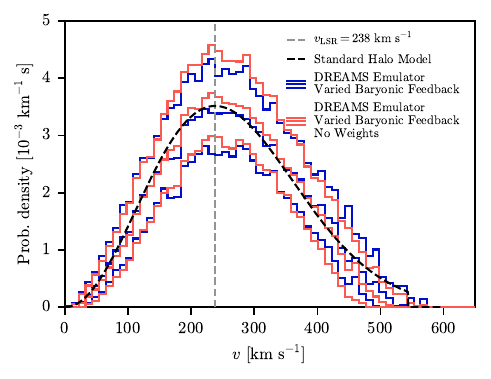}
  \caption{Similar to Figure~\ref{fig:milkyway}, except comparing the emulated DREAMS Varied dataset with~(dark blue) and without~(peach) the weighting scheme of Section~\ref{sub:weights}.  The weighting scheme slightly increases the distribution towards the highest speeds, but overall the results are insensitive to the choice of weights. }
  \label{fig:weights}
\end{figure*}

\begin{figure*}[h]
\centering
\includegraphics{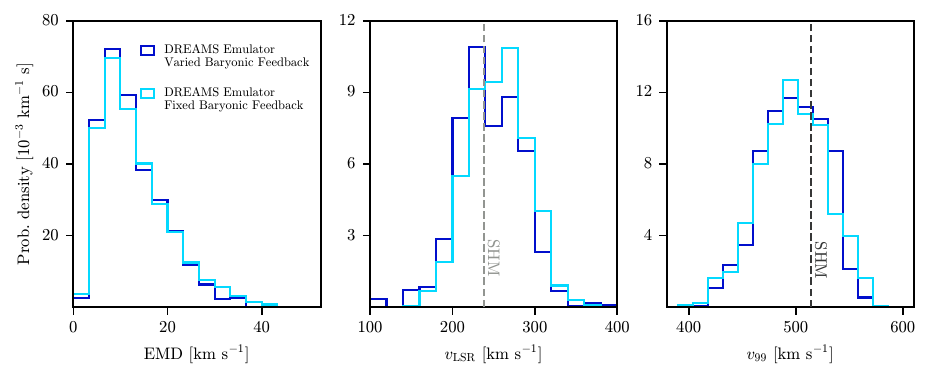}
  \caption{Distribution of EMD values (between the DREAMS result and SHM), peak speeds $v_{\rm LSR}$, and 99th-percentile speeds $v_{99}$.  The results are shown for the emulated DREAMS Varied dataset~(dark blue) and the emulated DREAMS Fixed dataset~(light blue). The SHM expectation is shown by the vertical dashed-gray line in the middle and right panel. }
  \label{fig:emd}
\end{figure*}

\begin{figure*}
\centering
\includegraphics{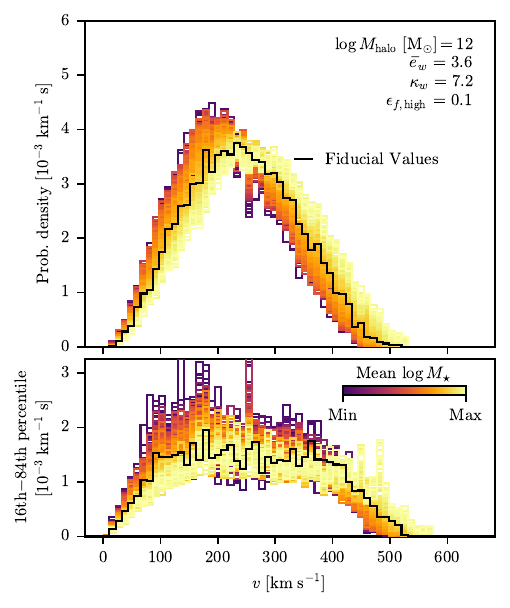}
  \caption{Same as Figure~\ref{fig:snfeedback}, except binned in terms of the logarithm of the host stellar mass, $\log M_*$.  
We consider 100 emulated samples each at 400 values of stellar mass. The stellar mass used is the mean stellar mass across the 100 emulated samples as we cannot fix its value given that it is an output parameter of the emulator.
As the stellar mass increases, the speed distributions shift to higher values.  The size of the shift is roughly comparable to that observed for \ew and $\kw$.  This is likely because supernova feedback is correlated with stellar mass (stronger feedback leads to less stellar mass).}
  \label{fig:mstellar}
\end{figure*}

\clearpage

\bibliographystyle{aasjournal}
\bibliography{references}



\label{lastpage}

\end{document}